\documentclass[10pt]{iopart}
\pdfoutput=1
\usepackage{graphicx}

\begin{document}

\title[Analytic Theory of ELM Suppression by Static RMPs]
{Analytic Theory of Edge Localized Mode Suppression by Static Resonant Magnetic Perturbations in H-mode Tokamak Discharges}

\author{Richard Fitzpatrick}

\address{Institute for Fusion Studies,  Department of Physics,  University of Texas at Austin, 
Austin TX, 78712, USA}

\ead{rfitzp@farside.ph.utexas.edu}

\begin{abstract}
  An analytic theory of edge localized mode (ELM) suppression in an H-mode tokamak plasma via the application of a static, externally generated, resonant magnetic perturbation (RMP) is presented. This theory is based on the plausible hypothesis that mode penetration at the top of the
  pedestal is a necessary and sufficient condition for the RMP-induced suppression of ELMs. 
  The theory also makes use of a number of  key insights gained in a recent publication (Fitzpatrick R 2019). The first insight is that the response of the plasma to a particular helical component of the RMP, in the immediate vicinity of the associated resonant surface, is governed by nonlinear magnetic island physics, rather than by linear layer physics. The second insight is that neoclassical effects play a vital role in the physics of RMP-induced ELM suppression.  The final insight is that plasma impurities play an important role in the physics of RMP-induced ELM suppression.  The theory presented  in this paper is employed 
 to gain a better understanding ELM suppression in DIII-D and ITER H-mode discharges.  It is found that ELM suppression
 is only possible when $q_{95}$  takes values that lie in certain narrow windows. Moreover, the widths of these windows decrease with increasing
 plasma density. Assuming a core plasma rotation of 20 krad/s, the window width for a model ITER H-mode discharge is found to be similar to, but slightly smaller than, the window width in a typical DIII-D
 H-mode discharge. 
\end{abstract}
\maketitle

\section{Introduction}
\subsection{Motivation for RMP-Generated ELM Suppression}
Tokamak discharges operating in high-confinement mode (H-mode) \cite{wagner} exhibit intermittent bursts of heat and particle transport, 
emanating from the outer regions of the plasma, that are known as {\em type-I edge localized modes} (ELMs) \cite{zohm}. 
ELMs are fairly harmless in present-day tokamaks possesing carbon plasma-facing components. However, large ELMs can cause
a problematic influx of tungsten ions into the plasma core in  tokamaks possessing tungsten plasma-facing components \cite{den}. 
Moreover, it is estimated that the heat load that ELMs
will deliver to the tungsten plasma-facing components in a reactor-scale tokamak, such as ITER, will be large enough to cause
massive tungsten ion influx into the core, and that the erosion associated with this process will 
unacceptably limit the lifetimes of these components \cite{loarte}. Consequently, developing robust and effective
methods for ELM control is a high priority for the international magnetic fusion program. 

The most promising method for the control of ELMs in H-mode tokamak discharges is via the application of static  {\em resonant magnetic perturbations}\/  (RMPs). Complete RMP-induced ELM suppression was first demonstrated on the DIII-D tokamak \cite{evans}. Subsequently, either mitigation or compete suppression of
ELMs has been demonstrated on the JET \cite{jet}, ASDEX-U \cite{asdex}, KSTAR \cite{kstar}, and EAST \cite{east} tokamaks.

\subsection{Observed Features of RMP-Generated ELM Suppression}
ELMs are  thought to
be caused by peeling-ballooning instabilities, with intermediate toroidal mode numbers, that are
driven by the strong pressure  and current density gradients characteristic of the edge region of an H-mode tokamak discharge \cite{conner}, which is generally known as the {\em pedestal}\/ region.
The initial observations of RMP-induced ELM suppression were interpreted as an indication that the magnetic field in the pedestal 
is  rendered stochastic by an applied RMP, leading to greatly enhanced transport via thermal diffusion along
magnetic field-lines \cite{evans,fenstermacher}. This explanation was quickly abandoned because no significant 
reduction in the electron temperature gradient in the pedestal is observed during RMP-induced  ELM suppression experiments,
whereas a very significant reduction would be expected in the presence of stochastic fields. It is now generally accepted that
response currents generated within the pedestal, as a consequence of plasma rotation,  play a crucial role in the perturbed equilibrium in the presence of RMPs, and that these currents  act to prevent the formation of RMP-driven magnetic island chains---a process known as {\em shielding}---and, thereby, significantly reduce the stochasticity of
the magnetic field \cite{berc}. 

The application of a static RMP, resonant in the pedestal region, to an H-mode tokamak discharge is observed to give rise to  two distinct phenomena \cite{schmitz, lanctot,paz1,d158115,paz}. The first of these  is the so-called {\em density pump-out}, which  is characterized by a reduction in the electron number density
in the pedestal region that varies smoothly with the amplitude of the applied RMP,  is accompanied by a similar, but significantly smaller, reduction
in the electron and ion temperatures, but  is not associated with ELM suppression. The second phenomenon is  {\em mode penetration}, which 
 occurs when the amplitude of the applied RMP exceeds a certain threshold value, is accompanied by sudden changes in the electron number
 density, electron and ion temperature, and ion toroidal  angular velocity,  profiles in the pedestal region, and, most importantly, is associated with the suppression of ELMs. 
Mode penetration is only observed to take place when $q_{95}$ (i.e., the safety-factor on the magnetic flux-surface that encloses 95\% of the poloidal flux enclosed by
the last closed flux-surface) takes values that lie in certain narrow windows \cite{paz1,d158115}. Furthermore, the mode penetration threshold is observed to increase markedly with
increasing electron number density \cite{paz1,d158115}.

\subsection{Recent Advances  in Theory of RMP-Generated ELM Suppression}\label{sadvance}
Recent research  \cite{hu} has shed considerable light on the hitherto poorly understood  physical mechanism that underlies RMP-induced ELM
suppression in H-mode tokamak discharges. In particular, computer simulations (made using the cylindrical, multi-harmonic, five-field, nonlinear, initial-value code, TM1 \cite{tm1,tm2,tm3}) of  RMP-induced ELM suppression experiments performed on the DIII-D tokamak  conclude that the density  pump-out  phenomenon is associated with the formation of a locked magnetic
island chain at the bottom of the pedestal, whereas the mode penetration phenomenon is associated with the formation of a locked magnetic island chain at the
top of the pedestal. 

As described below, the markedly different responses of the plasma to helical components of the applied RMP that are resonant at the bottom and at the
top of the pedestal---which is the main finding of the simulations described in \cite{hu}---can be accounted for on the basis of standard mode locking theory \cite{rf1,rf2}.  

The plasma at the bottom of the pedestal is too cold and
resistive for plasma rotation to effectively shield any locally resonant components of the applied RMP \cite{paper1}. Hence, if one of the helical harmonics of the RMP is resonant close to the bottom of the pedestal
then it generates a locked magnetic island chain whose width is similar to the associated vacuum island chain (i.e.,  the
island chain produced by naively superimposing the vacuum RMP onto the equilibrium plasma magnetic field). This island chain causes a local flattening of the
density and temperature profiles \cite{rf2a}. However, because the equilibrium density gradient at the bottom of the pedestal invariably greatly exceeds the
temperature gradient (see Figures~\ref{figb} and \ref{figf} and \cite{paper1}), this flattening produces a much more marked reduction in the pedestal density than in the pedestal temperature (i.e., the
flattening is responsible for the density pump-out phenomenon) \cite{hu}.

On the other hand, the plasma at the top of the pedestal is sufficiently hot that plasma rotation is capable of strongly shielding any locally resonant  components of the
applied RMP. Hence, if one of the helical harmonics of the RMP is resonant close to the top of the pedestal then it generates an island chain whose width
is much smaller than the associated vacuum island chain, and, consequently, has little effect on the local density and temperature profiles. 
However, if the amplitude of the resonant harmonic exceeds a certain critical value then torque balance at the resonant surface breaks down, resulting in the  sudden arrest of the
local plasma rotation, and the complete breakdown of shielding \cite{rf1,rf2}. This so-called mode penetration phenomenon is associated with the generation of a locked magnetic island chain at
the resonant surface whose width is similar to the vacuum island width. Such an island chain produces a local flattening of the plasma density and temperature profiles that is presumably sufficiently strong to prevent the plasma in the
pedestal region from ever exceeding the peeling-ballooning stability threshold. In fact, the fundamental hypothesis of both \cite{hu} and this paper is that
mode penetration at the top of the pedestal is directly responsible for ELM suppression. This hypothesis  allows us to model ELM suppression
by modeling mode penetration, which is a comparatively straightforward task,  rather than by directly modeling ELMs, which is a very much more complicated task.

\subsection{Further Advances  in Theory of RMP-Generated ELM Suppression}
Recent research \cite{paper1}, motivated by \cite{hu}, has lead to three key insights regarding the physics of mode penetration in the pedestal region of an
H-mode tokamak discharge. 

The first insight is that, in the immediate vicinity of
the associated resonant surface,  the response of the pedestal plasma to a particular helical component of the applied RMP is governed by {\em nonlinear}\/ magnetic island physics, rather than by {\em linear}\/ layer physics. The key point is that, although
plasma rotation at the top of the pedestal is capable of strongly shielding any locally resonant component of the RMP, this shielding is not strong enough
to reduce the width of the driven magnetic island chain below the linear layer width (which is very small in a high temperature tokamak plasma). The fact that
the shielded state is governed by nonlinear island physics, rather than by linear layer physics, has two important consequences. First, the shielded state consists of a narrow, {\em rotating}, magnetic island chain whose width
{\em pulsates}, rather than the constant-width locked island chain (whose width must, of course, be less than the linear layer width) predicted by linear physics
\cite{rf2,rfx,rfy,rf3}. (Incidentally, pulsating islands have been observed experimentally---see Figure 29 of \cite{raffi}.)  Second, the so-called {\em natural phase velocity}\/ of the driven
island chain (which is defined as the helical phase velocity  of the corresponding naturally
unstable island chain in the absence of the RMP \cite{rf1}) predicted by nonlinear island theory is
quite different to that predicted by linear layer theory. This is significant because both  the degree of shielding
at the resonant surface and the mode penetration threshold depend crucially on the natural phase velocity \cite{rf1}. According to linear layer theory, a naturally unstable magnetic island chain propagates in the
{\em electron}\/ diamagnetic direction relative to the ${\bf E}\times{\bf B}$ frame at the resonant surface \cite{ara}.
On the other hand, according to nonlinear island theory, a  naturally unstable magnetic island chain propagates in the
{\em ion}\/ diamagnetic direction relative to the  local ${\bf E}\times{\bf B}$ frame \cite{rfy,rf3}. Incidentally, there is clear experimental confirmation that a naturally unstable nonlinear magnetic island chain 
in a tokamak plasma propagates in the ion, rather than the electron, diamagnetic direction relative to the
local ${\bf E}\times{\bf B}$ frame \cite{lahaye,buratti}.

The second insight is that {\em neoclassical}\/ effects play a crucial role in  the physics of mode penetration in the
pedestal region of an
H-mode tokamak discharge. This is the case, firstly, because intrinsic neoclassical poloidal rotation is the main controlling
factor that determines the natural phase velocity of a nonlinear magnetic island chain (relative to the local ${\bf E}\times{\bf B}$ frame) \cite{rfy,rf3}, secondly, because 
neoclassical poloidal flow-damping plays a very significant role in determining the mode penetration threshold \cite{paper1}, and, finally, because the neoclassical modification of the plasma electrical conductivity is particularly
large in the outer regions of a typical H-mode tokamak discharge (because of the large fraction of trapped particles), and must, therefore, be taken into account in order to accurately calculate
the plasma response to an applied RMP at a resonant surface lying in the pedestal region. 

The final insight is that plasma {\em impurities}\/ play an important role in the physics of mode penetration in the
pedestal region of an
H-mode tokamak discharge. This is the case because the presence of experimentally observed levels of plasma impurities 
can  significantly modify the intrinsic neoclassical poloidal rotation, the neoclassical poloidal
flow-damping rate, and the classical plasma electrical conductivity, in the outer regions of an
H-mode discharge \cite{paper1}.

\subsection{Purpose of Paper}
The purpose of this paper is to improve the model of RMP-induced mode penetration in the
pedestal region of an H-mode tokamak plasma that was derived in \cite{paper1}, and then
to use the model to gain a better understanding of the physics of  ELM suppression by RMPs.  
The former goal is achieved, firstly, by fully incorporating the well-known moment-based neoclassical theory of Hirshman and Sigmar into
the model \cite{sigmar}, secondly, by allowing for non-trace amounts of plasma impurities, and, finally, by incorporating
some useful mode penetration results that were first derived in  \cite{rfx}. 

\section{Preliminary Analysis}\label{s2}
\subsection{Plasma Equilibrium}
Consider an axisymmetric tokamak plasma equilibrium whose magnetic and electric fields
take the forms \cite{sigmar}
\begin{eqnarray}
{\bf B} &=& I\,\nabla\phi +\nabla\phi\times\nabla\psi,\\[0.5ex]
{\bf E} &=&-\nabla{\mit\Phi}+E_\parallel\,{\bf b},
\end{eqnarray}
respectively, where ${\bf b}= {\bf B}/B$, $I=I(\psi)$, and ${\mit\Phi} = {\mit\Phi}(\psi)$.
Here, $\phi$ is the geometric toroidal angle, which implies that $|\nabla\phi|=1/R$, where $R$ is the
perpendicular distance from the toroidal symmetry axis. Furthermore, $\psi$ is the poloidal magnetic
flux (which is a convenient label for magnetic flux surfaces), whereas ${\mit\Phi}$ is the electric
scalar potential. 

It is helpful to define a ``straight'' poloidal angle, $\theta$, which increases by $2\pi$ radians for every poloidal
circuit around a magnetic flux surface, and which is such that
\begin{equation}
\nabla\psi\cdot\nabla\theta\times\nabla\phi \equiv {\bf B}\cdot\nabla\theta = \frac{I}{R^{\,2}\,q},
\end{equation}
where $q(\psi)$ is the so-called safety factor. It follows that
\begin{equation}\label{e4}
{\bf B}\cdot\nabla = \frac{I}{R^{\,2}}\left(\frac{\partial}{\partial\phi}+\frac{1}{q}\,\frac{\partial}{\partial\theta}\right).
\end{equation}
It is also helpful to define a ``geometric'' poloidal angle \cite{sigmar}, 
\begin{equation}
{\mit\Theta}= \left.2\pi\int_0^\theta\frac{d\theta'}{{\bf b}\cdot\nabla\theta'}\right/\oint\frac{d\theta'}{{\bf b}\cdot\nabla\theta'}.
\end{equation}
It is easily demonstrated that
\begin{equation}\label{e6}
{\bf b}\cdot\nabla{\mit\Theta}={\bf b}\cdot\nabla{\mit\Theta}(\psi).
\end{equation} 

\subsection{Useful Definitions}
The plasma is assumed to be made up of a number of distinct species.  
Let $m_a$, $e_a$, $n_a(\psi)$, $T_a(\psi)$, and $p_a=n_a\,T_a$ be the mass, electric charge, number density, temperature (in energy units),
and pressure, respectively, of species $a$. The species-$a$ thermal velocity takes the form
\begin{equation}
v_{T\,a} = \sqrt{\frac{2\,T_a}{m_a}}.
\end{equation}
It is helpful to define the species-$a$
collision time, $\tau_{aa}$, where \cite{sigmar}
\begin{equation}
\frac{1}{\tau_{aa}}= \frac{4}{3\sqrt{\pi}}\,\frac{4\pi\,n_a\,e_a^{\,4}\,\ln{\mit\Lambda}}{(4\pi\,\epsilon_0)^{\,2}\,m_a^{\,2}\,v_{T\,a}^{\,3}},
\end{equation}
and the Coulomb logarithm, $\ln{\mit\Lambda}$, is assumed to take the same large constant value (i.e., $\ln{\mit\Lambda}\simeq 17$), 
independent of species.

The flux-surface average operator, $\langle\cdots\rangle$,  is defined \cite{sigmar}
\begin{eqnarray}
\fl \langle A\rangle(\psi)= \left.\oint \frac{A(\psi,\theta)\,d\theta}{{\bf B}\cdot\nabla\theta}\right/\oint
\frac{d\theta}{{\bf B}\cdot\nabla\theta}
=
\left.\oint \frac{A(\psi,{\mit\Theta})\,d{\mit\Theta}}{B(\psi,{\mit\Theta})}\right/\oint
\frac{d{\mit\Theta}}{B(\psi,{\mit\Theta})},
\end{eqnarray}
where use has been made of (\ref{e6}).
Finally, the  transit frequency of species $a$ takes the form
\begin{equation}\label{e10}
\omega_{t\,a} = \frac{v_{T\,a}}{L_c},
\end{equation}
where \cite{sigmar}
\begin{eqnarray}\label{e11}
\fl L_c =\frac{\langle B^{\,2}\rangle^{\,2}}{\langle ({\bf b}\cdot\nabla B)^{\,2}\,\rangle\,
|\langle {\bf B}\cdot\nabla\theta\rangle|}\nonumber\\[0.5ex]
\phantom{=}\times\sum_{k>0} \frac{2}{k}\left[\langle \sin(k\,{\mit\Theta})\,({\bf b}\cdot\nabla \ln B)\rangle\,
\left\langle\,\sin(k\,{\mit\Theta})\,\frac{({\bf b}\cdot\nabla \ln B)}{B}\right\rangle\right]
\end{eqnarray}
is the flux-surface-averaged magnetic connection length.

In the banana collisionality regime, the fraction of circulating particles on a given magnetic flux-surface is written \cite{sigmar}
\begin{equation}
f_c=\frac{3}{4}\,\langle B^{\,2}\rangle\int_0^{1/B_{\rm max}}\frac{\lambda\,d\lambda}{\langle \sqrt{1-\lambda\,B}\rangle}.
\end{equation}
Finally, the dimensionless species-$a$ collisionality parameter takes the form \cite{sigmar}
\begin{equation}\label{e13}
\nu_{\ast\,a} = \frac{8}{3\pi}\,\frac{\langle B^{\,2}\rangle}{\langle ({\bf b}\cdot\nabla B)^{\,2}\rangle}\,\frac{g\,\omega_{t\,a}}{v_{T\,a}^{\,2}\,\tau_{aa}},
\end{equation}
where
\begin{equation}
g = \frac{1-f_c}{f_c}.
\end{equation}

\subsection{Model Magnetic Equilibrium}
In the paper, for the sake of simplicity, we shall adopt a model magnetic equilibrium characterized by
\begin{eqnarray}
R(r,{\mit\Theta})&=& R_0\,(1+\epsilon\,\cos{\mit\Theta}),\\[0.5ex]
B(r,{\mit\Theta})&= &\frac{B_0}{1+\epsilon\,\cos{\mit\Theta}}.
\end{eqnarray}
Here, $R_0$ and $B_0$ are the major radius and toroidal magnetic field-strength, respectively, on the magnetic
axis. Furthermore, $\epsilon=r/R_0$ where $r$ is a flux-surface label with the dimensions of length. The magnetic axis lies at $r=0$, whereas the plasma boundary is assumed to lie at $r=a$. 
Of course, $I=B_0\,R_0$ and $a/R_0<1$. Strictly speaking, our model is only appropriate to
a large aspect-ratio, low-$\beta$, tokamak equilibrium whose magnetic flux-surfaces possess 
circular poloidal cross-sections. 

It is easily demonstrated that
\begin{eqnarray}
|\nabla{\mit\Theta}| = \frac{1}{r},\label{e17}\\[0.5ex]
\tan\left(\frac{\theta}{2}\right)=\left(\frac{1-\epsilon}{1+\epsilon}\right)^{1/2}\tan\left(\frac{\mit\Theta}{2}\right),\\[0.5ex]
\langle B^{\,2}\rangle =\frac{B_0^{\,2}}{(1-\epsilon^{\,2})^{\,1/2}},\\[0.5ex]
\langle ({\bf b}\cdot\nabla B)^{\,2}\rangle = \frac{B_0^{\,2}\,\epsilon^{\,2}}{2\,R_0^{\,2}\,q^{\,2}\,(1-\epsilon^{\,2})^{\,5/2}},\\[0.5ex]
\langle {\bf B}\cdot\nabla \theta\rangle=\langle {\bf B}\cdot\nabla {\mit\Theta}\rangle = \frac{B_0}{R_0\,q\,(1-\epsilon^{\,2})^{\,1/2}}\label{e21},\\[0.5ex]
\langle \sin(k\,{\mit\Theta})\,({\bf b}\cdot\nabla \ln B)\rangle=-\frac{\epsilon\,\delta_{k1}}{2\,R_0\,q\,(1-\epsilon^{\,2})^{\,1/2}},\\[0.5ex]
\left\langle\,\sin{\mit\Theta}\,\frac{({\bf b}\cdot\nabla \ln B)}{B}\right\rangle=- \frac{\epsilon}{2\,R_0\,q\,B_0\,(1-\epsilon^{\,2})^{\,1/2}}.
\end{eqnarray}
Hence, (\ref{e10}), (\ref{e11}), and (\ref{e13}) yield
\begin{eqnarray}
\omega_{t\,a} &=& \frac{v_{T\,a}}{R_0\,|q|\,(1-\epsilon^{\,2})},\\[0.5ex]
\nu_{\ast\,a} &=& \frac{16}{3\pi}\frac{g}{\epsilon^{\,2}\,\omega_{t\,a}\,\tau_{aa}}.
\end{eqnarray}
Finally, it is well-known that \cite{grob1}
\begin{equation}
f_c \simeq 1-1.46\,\epsilon^{\,1/2}+0.46\,\epsilon^{\,3/2}.
\end{equation}

\section{Neoclassical Theory}
\subsection{Plasma Species}
In the following, we shall assume that the plasma consists of three species; namely,  electrons ($e$), majority ions ($i$), and impurity ions
($I$).  The charges of the three species are $e_e=-e$, $e_i= e$, and $e_I=Z_I\,e$, respectively, where
$e$ is the magnitude of the electron charge. Incidentally, in all of the calculations presented in this paper, the mean impurity ion charge number, $Z_I$,
for a given impurity species is determined as a function of the electron temperature and density using data obtained from the {\sc FLYCHK} code \cite{flychk}.
Of course, quasi-neutrality demands that
\begin{equation}
n_e= n_i+ Z_I\,n_I.
\end{equation}
It is helpful to define the effective ion charge number,
\begin{equation}
Z_{\rm eff} =\frac{n_i+Z_I^{\,2}\,n_I}{n_e}.
\end{equation}
It follows from the previous two equations that
\begin{eqnarray}\label{e29}
\frac{n_i}{n_e}&=& \frac{Z_I-Z_{\rm eff}}{Z_I-1},\\[0.5ex]
\frac{n_I}{n_e}&=&\frac{Z_{\rm eff}-1}{Z_I\,(Z_I-1)}.\label{e30}
\end{eqnarray}
It is also helpful to define
\begin{eqnarray}
\alpha& \equiv&\frac{n_I\,Z_I^{\,2}}{n_i}= \frac{Z_I\,(Z_{\rm eff}-1)}{Z_I-Z_{\rm eff}},\\[0.5ex]
Z_{{\rm eff}\,i}&\equiv&\frac{n_i}{n_e}= \frac{Z_I-Z_{\rm eff}}{Z_I-1},\\[0.5ex]
Z_{{\rm eff}\,I}&\equiv& \frac{n_I\,Z_I^{\,2}}{n_e} = \frac{Z_I\,(Z_{\rm eff}-1)}{Z_I-1}.
\end{eqnarray}
Note that $Z_{\rm eff} = Z_{{\rm eff}\,i}+ Z_{{\rm eff}\,I}$. 

\subsection{Ion Collisional Friction Matrices}
Let
\begin{equation}
x_{ab}=\frac{v_{T\,b}}{v_{T\,a}}.
\end{equation}
In the following, all quantities that are of order $(m_e/m_i)^{\,1/2}$, $(m_e/m_I)^{\,1/2}$, or smaller, are neglected with respect to unity. 
The  $2\times 2$ dimensionless ion collisional friction matrices, $F^{\,ii}$, $F^{\,iI}$, $F^{\,Ii}$, and $F^{\,II}$, are defined to have the following elements \cite{sigmar,grob1}:
\begin{eqnarray}
F^{\,ii}_{\,00} &=& \frac{\alpha\,(1+m_i/m_I)}{(1+x_{iI})^{\,3/2}},\\[0.5ex]
F^{\,ii}_{\,01}&=&\frac{3}{2}\,\frac{\alpha\,(1+m_i/m_I)}{(1+x_{iI}^{\,2})^{\,5/2}},\\[0.5ex]
F^{\,ii}_{\,10}&=&F^{\,ii}_{\,01},\\[0.5ex]
F^{\,ii}_{\,11}& =&\sqrt{2}+ \frac{\alpha\,[13/4+4\,x_{iI}^{\,2}+(15/2)\,x_{iI}^{\,4}]}{(1+x_{iI}^{\,2})^{\,5/2}},\\[0.5ex]
F^{\,iI}_{\,00} &=&F^{\,ii}_{\,00},\\[0.5ex]
F^{\,iI}_{\,01}&=& \frac{3}{2}\,\frac{T_i}{T_I}\,\frac{\alpha\,(1+m_I/m_i)}{x_{iI}\,(1+x_{Ii}^{\,2})^{\,5/2}},\\[0.5ex]
F^{\,iI}_{\,10}&=&F^{\,ii}_{01},\\[0.5ex]
F^{\,iI}_{\,11}& =&\frac{27}{4}\,\frac{T_i}{T_I}\,\frac{\alpha\,x_{iI}^{\,2}}{(1+x_{iI}^{\,2})^{\,5/2}},\\[0.5ex]
F^{\,Ii}_{\,00} &=&F^{\,ii}_{\,00},\\[0.5ex]
F^{\,Ii}_{\,01}&=&F^{\,ii}_{01},\\[0.5ex]
F^{\,Ii}_{\,10}&=&F^{\,iI}_{01},\\[0.5ex]
F^{\,Ii}_{\,11}& =&\frac{27}{4}\,\frac{\alpha\,x_{iI}^{\,2}}{(1+x_{iI}^{\,2})^{\,5/2}},\\[0.5ex]
F^{\,II}_{\,00} &=&F^{\,ii}_{00},\\[0.5ex]
F^{\,II}_{\,01}&=&F^{\,iI}_{01},\\[0.5ex]
F^{\,II}_{\,10}&=&F^{\,iI}_{01},\\[0.5ex]
F^{\,II}_{\,11}& =&\frac{T_i}{T_I}\left\{\sqrt{2}\,\alpha^{\,2}\,x_{Ii} + \frac{\alpha\,[
15/2+4\,x_{iI}^{\,2}+(13/4)\,x_{iI}^{\,4}]}{(1+x_{iI}^{\,2})^{\,5/2}}\right\}.
\end{eqnarray}
Note that $F^{\,ii}_{0j}=F^{\,Ii}_{0j}$, $F^{\,iI}_{0j}=F^{\,II}_{0j}$, $F^{\,ii}_{j0}=F^{\,iI}_{j0}$, and $F^{\,Ii}_{j0}=F^{\,II}_{j0}$, where
$j=0$, $1$. 

\subsection{Electron Collisional Friction Matrices}
The $2\times 2$ dimensionless electron collisional friction matrices, $F^{\,ee}$, $F^{\,ei}$, and $F^{\,eI}$,  are defined to have the following elements \cite{sigmar}:
\begin{eqnarray}
F^{\,ee}_{00} &= &Z_{\rm eff},\\[0.5ex]
F^{\,ee}_{01} &= &\frac{3}{2}\,Z_{\rm eff},\\[0.5ex]
F^{\,ee}_{10} &= &\frac{3}{2}\,Z_{\rm eff},\\[0.5ex]
F^{\,ee}_{11} &= &\sqrt{2} + \frac{13}{4}\,Z_{\rm eff},\\[0.5ex]
F^{\,ei}_{00} &= &Z_{{\rm eff}\,i},\\[0.5ex]
F^{\,ei}_{01} &=& 0,\\[0.5ex]
F^{\,ei}_{10} &= &\frac{3}{2}\,Z_{{\rm eff}\,i},\\[0.5ex]
F^{\,ei}_{11} &=& 0,\\[0.5ex]
F^{\,eI}_{00} &=& Z_{{\rm eff}\,I},\\[0.5ex]
F^{\,eI}_{01} &=& 0,\\[0.5ex]
F^{\,eI}_{10} &= &\frac{3}{2}\,Z_{{\rm eff}\,I},\\[0.5ex]
F^{\,eI}_{11} &= &0.
\end{eqnarray}
Note that $F^{\,ee}_{j0}= F^{\,ei}_{j0} + F^{\,eI}_{j0}$, where $j=0$, $1$. 

\subsection{Neoclassical Viscosity Matrices}\label{vmatrix}
The $2\times 2$ dimensionless species-$a$ neoclassical viscosity matrix, $\mu^{\,a}$,  is defined to have the following elements \cite{sigmar}: 
\begin{eqnarray}
\mu_{00}^{\,a} &=& K_{00}^{\,a},\\[0.5ex]
\mu_{01}^{\,a}&=& \frac{5}{2}\,K_{00}^{\,a}- K_{01}^{\,a},\\[0.5ex]
\mu_{10}^{\,a}&= &\mu_{01}^{\,a},\\[0.5ex]
\mu_{11}^{\,a} &=& K_{11}^{\,a} - 5\,K_{01}^{\,a}+\frac{25}{4}\,K_{00}^{\,a}.
\end{eqnarray}
Here,
\begin{eqnarray}
\fl K_{ab}^{\,e} = g\,\frac{4}{3\sqrt{\pi}}\int_0^\infty
\frac{e^{-x}\,x^{\,4+a+b}\,\nu_D^{\,e}(x)\,dx}{[x^{\,2}+\nu_{\ast\,e}\,\nu_D^{\,e}(x)]\,[x^{\,2}+(5\pi/8)\,(\omega_{t\,e}\,\tau_{ee})^{\,-1}\,\nu_T^{\,e}(x)]},\nonumber\\[0.5ex]\\[0.5ex]
\nu_D^{\,e}= \frac{3\sqrt{\pi}}{4}\left[\left(1-\frac{1}{2\,x}\right)\psi(x)+\psi'(x)\right]+\frac{3\sqrt{\pi}}{4}\,Z_{\rm eff},\\[0.5ex]
\nu_\epsilon^{\,e}= \frac{3\sqrt{\pi}}{2}\left[\psi(x)-\psi'(x)\right],\\[0.5ex]
\nu_T^{\,e}(x) = 3\,\nu_D^{\,e}(x)+\nu_{\epsilon}^{\,e}(x),
\end{eqnarray}
and
\begin{eqnarray}
\psi(x) &=& \frac{2}{\sqrt{\pi}}\int_0^x{\rm e}^{-t^{\,2}}\,dt -\frac{2}{\sqrt{\pi}}\,x\,{\rm e}^{-x^{\,2}},\\[0.5ex]
\psi'(x)&= &\frac{2}{\sqrt{\pi}}\,x\,{\rm e}^{-x^{\,2}}.
\end{eqnarray}
Furthermore, 
\begin{eqnarray}
\fl K_{ab}^{\,i} =g\,\frac{4}{3\sqrt{\pi}}\int_0^\infty
\frac{e^{-x}\,x^{\,2+a+b}\,\nu_D^{\,i}(x)\,dx}{[x+\nu_{\ast\,i}\,\nu_D^{\,i}(x)]\,[x+(5\pi/8)\,(\omega_{t\,i}\,\tau_{ii})^{\,-1}\,\nu_T^{\,i}(x)]},\\[0.5ex]
\fl \nu_D^{\,i}= \frac{3\sqrt{\pi}}{4}\left[\left(1-\frac{1}{2\,x}\right)\psi(x)+\psi'(x)\right]\frac{1}{x}\nonumber\\[0.5ex]\phantom{===}
+\frac{3\sqrt{\pi}}{4}\,\alpha\left[\left(1-\frac{x_{iI}}{2\,x}\right)\psi\!\left(\frac{x}{x_{iI}}\right)
+\psi'\!\left(\frac{x}{x_{iI}}\right)\right]\frac{1}{x},\\[0.5ex]
\fl \nu_\epsilon^{\,i}=\frac{3\sqrt{\pi}}{2}\left[\psi(x)-\psi'(x)\right]\frac{1}{x}\nonumber\\[0.5ex]\phantom{===}
+\frac{3\sqrt{\pi}}{2}\,\alpha\left[\frac{m_i}{m_I}\,\psi\!\left(\frac{x}{x_{iI}}\right)
-\psi'\!\left(\frac{x}{x_{iI}}\right)\right]\frac{1}{x},\nonumber\\[0.5ex]
\nu_T^{\,i}(x) = 3\,\nu_D^{\,i}(x)+\nu_{\epsilon}^{\,i}(x),
\end{eqnarray}
and, finally, 
\begin{eqnarray}
\fl K_{ab}^{\,I} = g\,\frac{4}{3\sqrt{\pi}}\int_0^\infty
\frac{e^{-x}\,x^{\,2+a+b}\,\nu_D^{\,I}(x)\,dx}{[x+\nu_{\ast\,I}\,\nu_D^{\,I}(x)]\,[x+(5\pi/8)\,(\omega_{t\,I}\,\tau_{II})^{\,-1}\,\nu_T^{\,I}(x)]},\\[0.5ex]
\fl \nu_D^{\,I}= \frac{3\sqrt{\pi}}{4}\left[\left(1-\frac{1}{2\,x}\right)\psi(x)+\psi'(x)\right]\frac{1}{x}\nonumber\\[0.5ex]\phantom{===}
+\frac{3\sqrt{\pi}}{4}\,\frac{1}{\alpha}\left[\left(1-\frac{x_{Ii}}{2\,x}\right)\psi\!\left(\frac{x}{x_{Ii}}\right)
+\psi'\!\left(\frac{x}{x_{Ii}}\right)\right]\frac{1}{x},\\[0.5ex]
\fl \nu_\epsilon^{\,I}= \frac{3\sqrt{\pi}}{2}\left[\psi(x)-\psi'(x)\right]\frac{1}{x}\nonumber\\[0.5ex]\phantom{===}
+\frac{3\sqrt{\pi}}{2}\,\frac{1}{\alpha}\left[\frac{m_I}{m_i}\,\psi\!\left(\frac{x}{x_{Ii}}\right)
-\psi'\!\left(\frac{x}{x_{Ii}}\right)\right]\frac{1}{x},\\[0.5ex]
\nu_T^{\,I}(x) = 3\,\nu_D^{\,I}(x)+\nu_{\epsilon}^{\,I}(x).
\end{eqnarray}
Note that our expressions for the neoclassical viscosity matrices interpolate (in the most accurate manner possible) between the
three standard neoclassical collisionality  regimes (i.e., the banana, plateau, and Pfirsch-Schl\"uter regimes \cite{sigmar}).

\subsection{Parallel Force and Heat Balance}\label{sbalance}
Let
\begin{equation}
\tilde{\mu}^{\,I} =\alpha^{\,2}\,\frac{T_i}{T_I}\,x_{Ii}\,\mu^{\,I}.
\end{equation}
The requirement of equilibrium force and heat balance parallel to the magnetic field \cite{sigmar,grob1} leads us to define
four $2\times 2$  dimensionless ion matrices, $L^{\,ii}(\psi)$, $L^{\,iI}(\psi)$, $L^{\,Ii}(\psi)$, and $L^{\,II}(\psi)$,
where,
\begin{eqnarray}
\fl\left(\begin{array}{cc} L^{\,ii}, & L^{\,iI}\\[0.5ex] L^{\,Ii},& L^{\,II}\end{array}\right)&=&
\left(\begin{array}{cc} F^{\,ii}+\mu^{\,i}, & -F^{\,iI}\\[0.5ex] -F^{\,Ii}, & F^{\,II}+\tilde{\mu}^{\,I}\end{array}\right)^{-1}
\left(\begin{array}{cc} F^{\,ii}, & -F^{\,iI}\\[0.5ex] -F^{\,Ii}, & F^{\,II}\end{array}\right).
\end{eqnarray}
and four $2\times 2$ dimensionless electron matrices, $Q^{\,ee}(\psi)$, $L^{\,ee}(\psi)$, $L^{\,ei}(\psi)$, and $L^{\,eI}(\psi)$, 
where
\begin{eqnarray}
Q^{\,ee}&=& (F^{\,ee}+\mu^{\,e})^{\,-1},\\[0.5ex]
L^{\,ee} &=& Q^{\,ee}\,F^{\,ee},\\[0.5ex]
L^{\,ei} &= &-Q^{\,ee}\left[F^{\,ei}\,(1-L^{\,ii})-F^{\,eI}\,L^{\,Ii}\right],\\[0.5ex]
L^{\,eI} &=& -Q^{\,ee}\left[F^{\,eI}\,(1-L^{\,II})-F^{\,ei}\,L^{\,iI}\right].
\end{eqnarray}

\subsection{Neoclassical Ion Velocities}
It is helpful to define the ${\bf E}\times {\bf B}$ frequency,
\begin{equation}
\omega_E(\psi) =-\frac{d{\mit\Phi}}{d\psi},
\end{equation}
the species-$a$ diamagnetic frequency,
\begin{equation}
\omega_{\ast\,a}(\psi) = -\frac{T_a}{e_a}\,\frac{d\ln p_a}{d\psi},
\end{equation}
and
\begin{equation}
\eta_a(\psi) = \frac{d\ln T_a}{d\ln n_a}.
\end{equation}
Let ${\bf V}^{\,a}$ denote the fluid velocity of species-$a$. 
The equilibrium neoclassical poloidal and parallel fluid velocities of the two ion species are \cite{sigmar,grob1}:
\begin{eqnarray}
\fl \frac{{\bf V}^{\,i}\cdot\nabla\theta}{{\bf B}\cdot\nabla\theta}\,\frac{\langle B^{\,2}\rangle}{I} = -L^{\,ii}_{00}\,\omega_{\ast\,i}+L^{\,ii}_{01}\left(\frac{\eta_i}{1+\eta_i}\right)\omega_{\ast\,i}-
L^{\,iI}_{00}\,\omega_{\ast\,I}
+L^{\,iI}_{01}\left(\frac{\eta_I}{1+\eta_I}\right)\omega_{\ast\,I},\\[0.5ex]
\fl \frac{{\bf V}^{\,I}\cdot\nabla\theta}{{\bf B}\cdot\nabla\theta}\,\frac{\langle B^{\,2}\rangle}{I} =-
L^{\,Ii}_{00}\,\omega_{\ast\,i}+L^{\,Ii}_{01}\left(\frac{\eta_i}{1+\eta_i}\right)\omega_{\ast\,i} -L^{\,II}_{00}\,\omega_{\ast\,I}
+L^{\,II}_{01}\left(\frac{\eta_I}{1+\eta_I}\right)\omega_{\ast\,I},\\[0.5ex]
\fl \frac{\langle {\bf V}^{\,i}\cdot {\bf B}\rangle}{I} =\omega_E +\omega_{\ast\,i}-L^{\,ii}_{00}\,\omega_{\ast\,i}+L^{\,ii}_{01}\left(\frac{\eta_i}{1+\eta_i}\right)\omega_{\ast\,i}-
L^{\,iI}_{00}\,\omega_{\ast\,I}
\nonumber\\[0.5ex]\phantom{======}
+L^{\,iI}_{01}\left(\frac{\eta_I}{1+\eta_I}\right)\omega_{\ast\,I},\label{e92}\\[0.5ex]
\fl \frac{\langle {\bf V}^{\,I} \cdot{\bf B}\rangle}{I} = \omega_E +\omega_{\ast\,I}-
L^{\,Ii}_{00}\,\omega_{\ast\,i}+L^{\,Ii}_{01}\left(\frac{\eta_i}{1+\eta_i}\right)\omega_{\ast\,i}-L^{\,II}_{00}\,\omega_{\ast\,I}
\nonumber\\[0.5ex]\phantom{======}
+L^{\,II}_{01}\left(\frac{\eta_I}{1+\eta_I}\right)\omega_{\ast\,I}.
\end{eqnarray}

\subsection{Parallel Current Density}
The equilibrium parallel current density can be written
\begin{equation}
j_\parallel = j_{\rm bootstrap} + j_{\rm ohmic},
\end{equation}
where \cite{sigmar,grob1}
\begin{eqnarray}
\fl \frac{\langle j_{\rm bootstrap}\,B\rangle}{I}=-\left[1-L^{\,ii}_{00} - \left(\frac{Z_{\rm eff}-1}{Z_I-Z_{\rm eff}}\right)L^{\,Ii}_{00}
+ \left(\frac{Z_I-1}{Z_I-Z_{\rm eff}}\right)L^{\,ei}_{00}\right]\frac{dp_i}{d\psi}\nonumber\\[0.5ex]
\phantom{=}- \left[L^{\,ii}_{01} + \left(\frac{Z_{\rm eff}-1}{Z_I-Z_{\rm eff}}\right)L^{\,Ii}_{01}
- \left(\frac{Z_I-1}{Z_I-Z_{\rm eff}}\right)L^{\,ei}_{01}\right]n_i\,\frac{dT_i}{d\psi}\nonumber\\[0.5ex]
\phantom{=}
- \left[1-L^{\,II}_{00} - \left(\frac{Z_I-Z_{\rm eff}}{Z_{\rm eff}-1}\right)L^{\,iI}_{00}
+ \left(\frac{Z_I-1}{Z_{\rm eff}-1}\right)L^{\,eI}_{00}\right]\frac{dp_I}{d\psi}
\nonumber\\[0.5ex]
\phantom{=}
- \left[L^{\,II}_{01} + \left(\frac{Z_I-Z_{\rm eff}}{Z_{\rm eff}-1}\right)L^{\,iI}_{01}
- \left(\frac{Z_I-1}{Z_{\rm eff}-1}\right)L^{\,eI}_{01}\right]n_I\,\frac{dT_I}{d\psi}
\nonumber\\[0.5ex]
\phantom{=}-(1-L^{\,ee}_{00})\,\frac{dp_e}{d\psi}-L^{\,ee}_{01}\,n_e\,\frac{dT_e}{d\psi}, 
\end{eqnarray}
and
\begin{equation}
\langle j_{\rm ohmic}\,B\rangle = Q_{00}^{\,ee}\,\sigma_{ee}\,\langle E_\parallel\,B\rangle,
\end{equation}
with
\begin{equation}
\sigma_{ee} = \frac{n_e\,e^{\,2}\,\tau_{ee}}{m_e}.
\end{equation}

\section{Plasma Response to Resonant Magnetic Perturbation}\label{s3}
\subsection{Plasma Response}
Consider the response of the plasma to a static RMP.  Suppose that the RMP
has $|m|$ periods in the poloidal direction, and
$n>0$ periods in the toroidal direction. (Note that $m$ is positive if $q$ is positive, and vice versa.) It is convenient to express the perturbed magnetic field in terms of the perturbed poloidal flux $\delta\psi(r,\theta,\phi,t)$:
\begin{equation}\label{e98}
\delta {\bf B} =\nabla\delta\psi\times R\,\nabla\phi,
\end{equation}
where 
\begin{equation}
\delta\psi(r,\theta,\phi,t)= \hat{\psi}(r,t)\,\exp[\,{\rm i}\,(m\,\theta-n\,\phi)].
\end{equation}

As is well known, the response of the plasma to the applied  RMP is governed by the equations of
perturbed, marginally-stable (i.e., $\partial/\partial t \equiv 0$), {\em ideal magnetohydrodynamics}\/ (MHD) everywhere in the plasma, apart
from a relatively narrow  (in $r$) region   in the vicinity of a  resonant magnetic flux-surface at which
${\bf B}\cdot\nabla\delta\psi = 0$ \cite{rf1}. 
It follows from (\ref{e4}) that the resonant surface is located at $r=r_s$, where $q(r_s)=m/n$ \cite{rf1}.

It is convenient to parameterize the RMP in terms of the so-called perturbed {\em vacuum magnetic flux}, ${\mit\Psi}_v(t)= |{\mit\Psi}_v|\,{\rm e}^{-{\rm i}\,\varphi_v}$,
which is defined to be the value of $\hat{\psi}(r,t)$ at  $r=r_s$ in the presence of the RMP, but in the absence of the plasma. Here,
$\varphi_v$ is the helical phase of the RMP, and is assumed to be constant in time.  Likewise, the response of the plasma in the vicinity of the
resonant surface to the RMP is parameterized in terms of the so-called {\em reconnected magnetic flux}, ${\mit\Psi}_s(t)= |{\mit\Psi}_s|\,{\rm e}^{-{\rm i}\,\varphi_s}$,
which is the actual value of $\hat{\psi}(r,t)$ at  $r=r_s$. Here, $\varphi_s(t)$ is the helical phase of the reconnected flux. 

The intrinsic stability of the $m$/$n$ tearing mode is governed by the {\em tearing stability index}\/ \cite{fkr}, 
\begin{equation}
{\mit\Delta}' = \left[\frac{d\ln\hat{\psi}}{dr}\right]_{r_s-}^{r_s+},
\end{equation}
where $\hat{\psi}(r)$ is a solution of the marginally-stable, ideal-MHD equations,  for the case of an $m$/$n$ helical perturbation, that satisfies physical boundary conditions at $r=0$ and $r=a$ (in the
absence of the RMP). According to resistive-MHD theory \cite{fkr,ruth}, if ${\mit\Delta}'>0$ then  the $m$/$n$ tearing mode spontaneously reconnects magnetic flux at the resonant
surface to form a helical magnetic island chain. In the following, it is assumed that ${\mit\Delta}'<0$, so that the $m$/$n$ tearing mode is intrinsically stable. In this
situation, any magnetic reconnection that takes place at the resonant surface is due solely to the RMP. 

\subsection{Nonlinear Response Regime}
In the {\em nonlinear}\/ local plasma response model adopted in this paper, the reconnected magnetic flux induced  at the resonant surface by the RMP is governed by two
equations. The first of these is the {\em Rutherford island width evolution equation}\/ \cite{ruth}, 
\begin{equation}
{\cal I}\,\tau_R\,\frac{d}{dt}\!\left(\frac{W}{r_s}\right) = {\mit\Delta}' \,r_s+ 2\,|m|\,{\cal A}\left(\frac{W_v}{W}\right)^2\cos\varphi,
\end{equation}
where ${\cal I}=0.8227$. Here,
\begin{equation}\label{e104}
\tau_R =\mu_0\left(r^{\,2}\,Q^{\,ee}_{00}\,\sigma_{ee}\right)_{r=r_s},
\end{equation}
is the resistive evolution timescale, and
\begin{equation}
W(t) = 4\left(\frac{|{\mit\Psi}_s|}{s\,r_s\,|B_p(r_s)|}\right)^{1/2}r_s
\end{equation}
 the full (radial) width of the magnetic island chain that forms at the resonant surface. (In this paper,
 it is implicitly assumed that $W\ll a$.) 
  Moreover, the neoclassical coefficient $Q^{\,ee}_{00}(r)$ is specified in Section~\ref{sbalance}, and 
 \begin{equation}\label{e105}
 B_p \equiv \frac{\langle {\bf B}\cdot\nabla{\mit\Theta}\rangle}{|\nabla{\mit\Theta}|}=  \frac{\epsilon\,B_0}{q\,\sqrt{1-\epsilon^{\,2}}}
 \end{equation}
 is the equilibrium poloidal magnetic field. [See (\ref{e17}) and (\ref{e21}).]
 Note that (\ref{e104}) fully takes into account the modification of plasma parallel electrical conductivity
 due to the presence of impurity ions, as well as the modification due to trapped particles. Moreover,
  (\ref{e104})  is
 valid in all three possible neoclassical collisionality regimes (i.e., banana, plateau, and Pfirsch-Schl\"uter \cite{sigmar}).
 Now,
\begin{equation}\label{e106}
W_v(t) = 4\left(\frac{|{\mit\Psi}_v|}{s\,r_s\,|B_p(r_s)|}\right)^{1/2}r_s
\end{equation}
is termed the vacuum island width. In addition,
$s=(d\ln q/d\ln r)_{r=r_s}$ is the local magnetic shear, and ${\cal A}$ the amplification factor (i.e., the factor by which the radial
magnetic field at the resonant surface due to the RMP is enhanced with respect to its vacuum value due to equilibrium plasma
currents external to the resonant surface). Finally,
\begin{equation}
\varphi(t) = \varphi_s(t)-\varphi_v
\end{equation}
is the helical phase of the island chain relative to the RMP. 
 
 The second governing equation is the so-called {\em no-slip constraint}\/ \cite{rf1}, 
\begin{equation}
\frac{d\varphi_s}{dt} - \omega=0,
\end{equation}
according to which the island chain is convected by the ``plasma'' (see the following sentence in parentheses) at the resonant surface. Here, 
\begin{equation}\label{e14}
\varpi(t) = m\,{\mit\Omega}_{\theta}(r_s,t) - n\,{\mit\Omega}_{\phi}(r_s,t)
\end{equation}
is the island phase velocity, and 
 ${\mit\Omega}_{\theta}(r,t)$ and ${\mit\Omega}_{\phi}(r,t)$ are the   plasma poloidal and toroidal angular velocity profiles, respectively. [To be more exact,  ${\mit\Omega}_{\theta}(r,t)$ and ${\mit\Omega}_{\phi}(r,t)$ are
the poloidal and toroidal angular velocity profiles of an imaginary fluid that convects reconnected magnetic flux
at resonant surfaces. It is assumed that changes in these velocity profiles are mirrored by changes in the corresponding actual
plasma velocity profiles.  A magnetic island convected by the imaginary fluid propagates at its so-called natural phase velocity. The relationship between the
natural phase velocity and the ${\bf E}\times {\bf B}$ frequency is specified in Section~\ref{snat}.] 

The nonlinear response model adopted in this paper is valid as long as the  width of the RMP-generated magnetic island
chain  at the resonant surface  exceeds the
linear layer width. This criterion is easily satisfied in present-day tokamaks, and will almost certainly
be satisfied in ITER \cite{paper1}. 

\subsection{Plasma Angular Velocity Evolution}\label{angular}
It is easily demonstrated that zero net electromagnetic torque can be exerted on magnetic flux surfaces located in a region of the
plasma that is governed by the equations of marginally-stable, ideal-MHD \cite{rf1}. Thus, any electromagnetic torque exerted on the plasma by the RMP
develops in the immediate vicinity of the resonant surface, where ideal-MHD breaks down. The
net  poloidal and toroidal  electromagnetic torques  exerted in the vicinity of the resonant surface by the RMP  take the forms \cite{rf1,rf2}
\begin{eqnarray}
T_{\theta\,{\rm EM}} &=& -\frac{4\pi^{\,2}\,|m|\,m\,R_0}{\mu_0}\,{\cal A}\,|{\mit\Psi}_v|\,|{\mit\Psi}_s|\,\sin\varphi,\\[0.5ex]
T_{\phi\,{\rm EM}} &=& \frac{4\pi^{\,2}\,|m|\,n\,R_0}{\mu_0}\,{\cal A}\,|{\mit\Psi}_v|\,|{\mit\Psi}_s|\,\sin\varphi,
\end{eqnarray}
respectively. 

We can write
\begin{eqnarray}\label{e22}
{\mit\Omega}_\theta(r,t) &=&{\mit\Omega}_{\theta\,0}(r) + {\mit\Delta\Omega}_\theta(r,t),\\[0.5ex]
{\mit\Omega}_\phi(r,t) &=&{\mit\Omega}_{\phi\,0}(r) + {\mit\Delta\Omega}_\phi(r,t),\label{e23}
\end{eqnarray}
where ${\mit\Omega}_{\theta\,0}(r)$ and ${\mit\Omega}_{\phi\,0}(r)$ are the equilibrium poloidal and toroidal
plasma angular velocity profiles, respectively, whereas ${\mit\Delta\Omega}_{\theta}(r,t)$ and ${\mit\Delta\Omega}_{\phi}(r,t)$ 
are the respective modifications to these profiles induced by the aforementioned electromagnetic torques. 
The modifications to the angular velocity profiles are governed by the poloidal and toroidal angular equations of
motion of the plasma, which take the respective forms\,\cite{rf1,hirsh}
\begin{eqnarray}\label{eom1}
\fl 4\pi^{\,2}\,R_0\left[(1+2\,q^{\,2})\,\rho\,r^{\,3}\,\frac{\partial{\mit\Delta\Omega}_\theta}{\partial t}-\frac{\partial}{\partial r}\!\left(
\mu\,r^{\,3}\,\frac{\partial{\mit\Delta\Omega}_\theta}{\partial r}\right) +\rho\,r^{\,3}\,\frac{{\mit\Delta\Omega}_\theta}{\tau_\theta}\right]\nonumber\\[0.5ex]\phantom{==}=T_{\theta\,{\rm EM}}\,\delta(r-r_s),\\[0.5ex]
\fl 4\pi^{\,2}\,R_0^{\,3}\left[\rho\,r\,\frac{\partial{\mit\Delta\Omega}_\phi}{\partial t}-\frac{\partial}{\partial r}\!\left(
\mu\,r\,\frac{\partial{\mit\Delta\Omega}_\phi}{\partial r}\right) +\rho\,r\,\frac{{\mit\Delta\Omega}_\phi}{\tau_\phi}\right]
=T_{\phi\,{\rm EM}}\,\delta(r-r_s),\label{eom2}
\end{eqnarray}
and are subject to the spatial boundary conditions \cite{rf1}
\begin{eqnarray}
\frac{\partial{\mit\Delta\Omega}_\theta(0,t)}{\partial r} &=&
\frac{\partial{\mit\Delta\Omega}_\phi(0,t)}{\partial r} =0,\\[0.5ex]
{\mit\Delta\Omega}_\theta(a,t)&=&{\mit\Delta\Omega}_\phi(a,t)=0.
\end{eqnarray}
Here, $\mu(r)$ is the anomalous plasma perpendicular ion viscosity (due to plasma turbulence),
whereas 
\begin{eqnarray}
\fl \rho(r)\simeq m_i\,n_i+m_I\,n_I
= \left\{m_i\left(\frac{Z_I-Z_{\rm eff}}{Z_I-1}\right)+m_I\left[\frac{Z_{\rm eff}-1}{Z_I\,(Z_I-1)}\right]\right\}n_e
\end{eqnarray}
 is the plasma mass density. [See (\ref{e29}) and (\ref{e30}).]
Furthermore
\begin{equation}
\frac{1}{\tau_\theta(r)} = \left(\frac{B_0}{B_p}\right)^{2}\,\frac{\mu_{00}^{\,i}}{\tau_{ii}}=\left(\frac{q\,\sqrt{1-\epsilon^{\,2}}}{\epsilon}\right)^{2}\,\frac{\mu_{00}^{\,i}}{\tau_{ii}}
\end{equation}
is  the neoclassical poloidal flow-damping rate \cite{grob1,stix}. [See (\ref{e105}).] Here, the
neoclassical coefficient $\mu_{00}^{\,i}(r)$ is specified
in Section~\ref{vmatrix}. Note that the previous expression is valid
in all possible neoclassical collisionality regimes (i.e., banana, plateau, and Pfirsch-Schl\"uter \cite{sigmar}), and also fully takes into
account the presence of impurity ions. 
Finally, $1/\tau_\phi(r)$ is  the neoclassical toroidal flow-damping rate. The neoclassical
toroidal flow-damping is assumed to be generated by non-resonant components of the applied
RMP \cite{shaing,cole}.  

The factor $(1+2\,q^{\,2})$ in (\ref{eom1}) derives from the fact that incompressible poloidal
flow has a poloidally-varying toroidal component that effectively increases the plasma mass being accelerated by the poloidal
flow-damping force \cite{hirsh}.

It turns out that, in the presence of  poloidal and toroidal flow-damping, the
modifications to the plasma poloidal and toroidal angular velocity profiles are radially localized in the vicinity of the resonant
surface \cite{cole}. Assuming that this is the case,  it is a good approximation to replace $q(r)$, $\rho(r)$, $\mu(r)$, $\tau_\theta(r)$, and $\tau_\phi(r)$ in
 (\ref{eom1}) and (\ref{eom2}) by their values  at the resonant surface. We are, nevertheless, assuming that the localization width greatly exceeds the
 width of the magnetic island chain.

Equations~(\ref{e14}), (\ref{e22}), and (\ref{e23}) imply that
\begin{equation}
\varpi(t)=\varpi_0 +m\,{\mit\Delta\Omega}_{\theta}(r_s,t)-n\,{\mit\Delta\Omega}_{\phi}(r_s,t),
\end{equation}
where 
\begin{equation}
\varpi_0 = m\,{\mit\Omega}_{\theta\,0}(r_s)-n\,{\mit\Omega}_{\phi\,0}(r_s)
\end{equation}
is the so-called {\em natural phase velocity}\/ of the $m$/$n$ tearing mode \cite{rf1}. In other words,
$\varpi_0$ is the helical phase velocity of a naturally unstable $m$/$n$ tearing mode in the
absence of the RMP.

\subsection{Natural Phase Velocity}\label{snat}
It turns out that the natural phase velocity of a nonlinear magnetic island chain in a conventional tokamak plasma is predominately determined by the neoclassical parallel ion viscous stress tensor. According to the analysis of \cite{rf3},
\begin{equation}
 \varpi_0 = -n\left\langle \frac{V_\phi^{\,i}}{R}\right\rangle_{r=r_s} \simeq -n\left(\frac{\langle {\bf V}^{\,i}\cdot{\bf B}\rangle}{I}\right)_{r=r_s}.
\end{equation}
In other words, the island chain is effectively convected by the toroidal component of the equilibrium neoclassical majority
ion flow at the resonant surface, but not by the poloidal component. It follows from (\ref{e92}) that
\begin{eqnarray}\label{e123}
\fl \varpi_0=-n\left(\omega_E +\left[1-L^{\,ii}_{00}+L^{\,ii}_{01}\left(\frac{\eta_i}{1+\eta_i}\right)\right]\omega_{\ast\,i}
\right.\nonumber\\[0.5ex]\left.
-
\left[L^{\,iI}_{00}-L^{\,iI}_{01}\left(\frac{\eta_I}{1+\eta_I}\right)\right]\omega_{\ast\,I}\right)_{r=r_s}.
\end{eqnarray}
Here, the neoclassical coefficients $L^{\,ii}_{00}(r)$, $L^{\,ii}_{01}(r)$, $L^{\,iI}_{00}(r)$, and $L^{\,iI}_{01}(r)$ are specified in 
Section~\ref{sbalance}. Note that the previous expression is valid
in all possible neoclassical collisionality regimes (i.e., banana, plateau, and Pfirsch-Schl\"uter \cite{sigmar}), and also fully takes into
account the presence of impurity ions. 

Incidentally, according to linear theory \cite{ara}, the natural phase velocity is
\begin{equation}
\varpi_{\perp\,e} =-n\,(\omega_E+\omega_{\ast\,e})_{r=r_s},
\end{equation}
which is equivalent to saying that the island chain is convected by the electron fluid. 

\subsection{Unnormalized Nonlinear Plasma Response Model}
Our nonlinear plasma response model reduces to the following
closed set of equations \cite{paper1}:
\begin{eqnarray}\label{ea}
{\cal I}\,\tau_R\,\frac{d}{dt}\!\left(\frac{W}{r_s}\right) = {\mit\Delta}' \,r_s+ 2\,|m|\,{\cal A}\left(\frac{W_v}{W}\right)^2\cos\varphi,\\[0.5ex]
\frac{d\varphi}{dt}=  \varpi,\\[0.5ex]
 \varpi = \varpi_0+m\,{\mit\Delta\Omega}_\theta(r_s,t)-n\,{\mit\Delta\Omega}_\phi(r_s,t),\\[0.5ex]
\fl \left[(1+2\,q_s^{\,2})\,\rho_s\,r^{\,3}\,\frac{\partial{\mit\Delta\Omega}_\theta}{\partial t}-\frac{\partial}{\partial r}\!\left(
\mu_s\,r^{\,3}\,\frac{\partial{\mit\Delta\Omega}_{\theta}}{\partial r}\right)+\rho_s\,r^{\,3}\,\frac{{\mit\Delta\Omega}_{\theta}}{\tau_{\theta\,s}}\right]\nonumber\\[0.5ex]
\phantom{==}=-\frac{|m|\,m}{\mu_0}\,{\cal A}\left(\frac{W_v}{4\,r_s}\right)^2\left(\frac{W}{4\,r_s}\right)^2 \left[s\,r_s\,B_p(r_s)\right]^{\,2}\,\sin\varphi\,\delta(r-r_s),\\[0.5ex]
\fl R_0^{\,2}\left[\rho_s\,r\,\frac{\partial{\mit\Delta\Omega}_\phi}{\partial t}-\frac{\partial}{\partial r}\!\left(
\mu_s\,r\,\frac{\partial{\mit\Delta\Omega}_\phi}{\partial r}\right) +\rho_s\,r\,\frac{{\mit\Delta\Omega}_\phi}{\tau_{\phi\,s}}\right]\nonumber\\[0.5ex]
\phantom{==}=\frac{|m|\,n}{\mu_0}\,{\cal A}\left(\frac{W_v}{4\,r_s}\right)^2\left(\frac{W}{4\,r_s}\right)^2\left[s\,r_s\,B_p(r_s)\right]^{\,2}\,\sin\varphi\,\delta(r-r_s),\\[0.5ex]
\frac{\partial{\mit\Delta\Omega}_\theta(0,t)}{\partial r} =
\frac{\partial{\mit\Delta\Omega}_\phi(0,t)}{\partial r} =0,\\[0.5ex]
{\mit\Delta\Omega}_\theta(a,t)={\mit\Delta\Omega}_\phi(a,t)=0.\label{ez}
\end{eqnarray}
Here, $q_s=q(r_s)=m/n$, $\rho_s=\rho(r_s)$, $\mu_s=\mu(r_s)$, $\tau_{\theta\,s}=\tau_\theta(r_s)$, and $\tau_{\phi\,s}=\tau_\phi(r_s)$. 

\subsection{Normalized Nonlinear Plasma Response Model}
For the sake of simplicity, we shall assume that the quantities ${\cal A}$ and ${\mit\Delta}'\,r_s$ take their
vacuum values, $1$ and $-2\,|m|$, respectively. (The former assumption is equivalent to assuming that
there is negligible equilibrium plasma current external to the resonant surface.) 
The hydromagnetic and momentum confinement timescales are defined
\begin{eqnarray}
\tau_H &=& \frac{R_0}{|B_0|}\,\frac{\sqrt{\mu_0\,\rho(r_s)\,(1-\epsilon_s^{\,2})}}{n\,s},\\[0.5ex]
\tau_M &= &\frac{a^{\,2}\,\rho_s}{\mu_s},
\end{eqnarray}
respectively, where $\epsilon_s=\epsilon(r_s)=r_s/R_0$. 
It is convenient to adopt the following normalization scheme: 
$\hat{t} = t/\tau_H$, 
$\hat{r} = r/a$, 
$\hat{r}_s= r_s/a$,
$\hat{\varpi}= \varpi\,\tau_H$, 
$\hat{\varpi}_0= \varpi_0\,\tau_H$, 
$\hat{\varpi}_\theta = - m\,{\mit\Delta\Omega}_\theta\,\tau_H$,
$\hat{\varpi}_\phi= n\,{\mit\Delta\Omega}_\phi\,\tau_H$,
$\hat{W} = W/W_v$,
$\hat{\tau}_M = \tau_M/\tau_H$,
$\hat{\tau}_\theta = \tau_{\theta\,s}/\tau_H$,
and
$\hat{\tau}_\phi= \tau_{\phi\,s}/\tau_H$.
The normalized versions of (\ref{ea})--(\ref{ez}) are:
\begin{eqnarray}\label{estart}
\frac{{\cal I}\,S\,W_v}{2\,|m|\,r_s}\,\frac{d\hat{W}}{d\hat{t}}=-1 + \frac{\cos\varphi}{\hat{W}^{\,2}},
\\[0.5ex]
\frac{d\varphi}{d\hat{t}}= \hat{\varpi},\label{estart1}\\[0.5ex]
 \hat{\varpi} = \hat{\varpi}_0-\hat{\varpi}_\theta(\hat{r}_s,\hat{t})-\hat{\varpi}_\phi(\hat{r}_s,\hat{t}),\label{estart2}\\[0.5ex]
\fl (1+2\,q_s^{\,2})\,\hat{r}^{\,3}\,\frac{\partial\hat{\varpi}_\theta}{\partial\hat{t}}-
\frac{1}{\hat{\tau}_M}\,\frac{\partial}{\partial\hat{r}}\!\left(\hat{r}^{\,3}\,\frac{\partial\hat{\varpi}_\theta}{\partial\hat{r}}\right)+\frac{1}{\hat{\tau}_\theta}\,\hat{r}^{\,3}\,\hat{\varpi}_\theta 
\nonumber\\[0.5ex]\phantom{==}
= 
|m|\left(\frac{W_v}{4\,a}\right)^{4}\,\hat{W^{\,2}}\,\sin\varphi\,\delta(\hat{r}-\hat{r}_s),\\[0.5ex]
\fl \hat{r}\,\frac{\partial\hat{\varpi}_\phi}{\partial\hat{t}}-
\frac{1}{\hat{\tau}_M}\,\frac{\partial}{\partial\hat{r}}\!\left(\hat{r}\,\frac{\partial\hat{\varpi}_\phi}{\partial\hat{r}}\right)+\frac{1}{\hat{\tau}_\phi}\,\hat{r}\,\hat{\varpi}_\phi
= 
\zeta\,|m|\left(\frac{W_v}{4\,a}\right)^{4}\,\hat{W^{\,2}}\,\sin\varphi\,\delta(\hat{r}-\hat{r}_s),\\[0.5ex]
\frac{\partial\hat{\varpi}_\theta(0,\hat{t})}{\partial \hat{r}} =
\frac{\partial \hat{\varpi}_\phi(0,\hat{t})}{\partial \hat{r}} =0,\\[0.5ex]
\hat{\varpi}_\theta(1,\hat{t})=\hat{\varpi}_\phi(1,\hat{t})=0,\label{eend}
\end{eqnarray}
where
\begin{eqnarray}
S&=&\frac{\tau_R}{\tau_H},\\[0.5ex]
\zeta&=& \left(\frac{a}{R_0\,q_s}\right)^{2}.
\end{eqnarray}

\section{Pulsating Magnetic Island Chains}\label{spul}

\subsection{Cycle-Averaged Nonlinear Plasma Response Model}
As is described in \cite{rf2} and \cite{rf3,rfx,rfy}, a non-stationary solution of (\ref{estart})--(\ref{eend})  (i.e., a solution in which the magnetic island chain driven at the resonant surface does not have a fixed phase relation with respect to the
RMP) is characterized by an island chain whose helical phase continually increases (assuming, for
the sake of definiteness, that $\hat{\varpi}>0$) in time, and
whose width {\em pulsates}, periodically falling to zero, at which times its helical
phase  jumps by $-\pi$ radians. Suppose that the island chain is born (at zero width) at helical phase $\varphi_0$, attains its maximum width at helical phase $\varphi_0+\pi/2$, 
and disappears  at helical phase $\varphi_0+\pi$, at which time its helical phase jumps back to
$\varphi_0$, and the cycle repeats. It is helpful
to define the cycle-average operator \cite{rfx}:
\begin{equation}
\langle \cdots\rangle_{\varphi} = \frac{1}{\pi}\int_{\varphi_0}^{\varphi_0+\pi}(\cdots)\,d\varphi.
\end{equation}

Equations~(\ref{estart}) and (\ref{estart1}) can be combined to give 
\begin{equation}\label{e143}
\lambda\,\frac{du}{d\varphi} \simeq \cos\varphi - u^{\,2/3},
\end{equation}
where
\begin{eqnarray}
u&=&\hat{W}^{\,3},\\[0.5ex]
\lambda &=& \frac{\langle\hat{\varpi}\rangle_\varphi\,{\cal I}\,S\,W_v}{6\,|m|\,r_s}.\label{e145}
\end{eqnarray}
Let
\begin{equation}
T \equiv \hat{W}^{\,2}\,\sin\varphi= u^{\,2/3}\,\sin\varphi.
\end{equation}
According to the analysis of \cite{rfx}, (\ref{e143}) possesses pulsating solutions
characterized by
\begin{eqnarray}
\frac{\langle\varphi\rangle_\varphi}{\pi}&\simeq &\frac{\lambda^{\,2}}{1+2\,\lambda^{\,2}}-\frac{0.1756\,\lambda^{\,2}}{1+0.4950\,|\lambda|^{\,8/3}},\\[0.5ex]
\langle\hat{W}^{\,4}\rangle_\varphi& \simeq &\frac{0.5}{1+0.8624\,|\lambda|^{\,4/3}},\\[0.5ex]
\langle T\rangle_\varphi &\simeq& \frac{0.4577\,\lambda}{1+0.8546\,|\lambda|^{\,5/3}}.\label{e149}
\end{eqnarray}

Finally, the cycle average of (\ref{estart2})--(\ref{eend}) yields
\begin{eqnarray} 
\langle\hat{\varpi}\rangle_\varphi = \hat{\varpi}_0-\langle\hat{\varpi}_\theta\rangle_\varphi(\hat{r}_s)-\langle\hat{\varpi}_\phi\rangle_\varphi(\hat{r}_s),\label{e150}\\[0.5ex]
\fl -\frac{1}{\hat{\tau}_M}\,\frac{d}{d\hat{r}}\!\left(\hat{r}^{\,3}\,\frac{d\langle\hat{\varpi}_\theta\rangle_\varphi}{d\hat{r}}\right)+\frac{1}{\hat{\tau}_\theta}\,\hat{r}^{\,3}\,\langle\hat{\varpi}_\theta\rangle_\varphi
= 
|m|\left(\frac{W_v}{4\,a}\right)^{4}\,\langle T\rangle_\varphi\,\delta(\hat{r}-\hat{r}_s),\\[0.5ex]
\fl -\frac{1}{\hat{\tau}_M}\,\frac{d}{d\hat{r}}\!\left(\hat{r}\,\frac{d\langle\hat{\varpi}_\phi\rangle_\varphi}{d\hat{r}}\right)+\frac{1}{\hat{\tau}_\phi}\,\hat{r}\,\langle\hat{\varpi}_\phi\rangle_\varphi
 = 
\zeta\,|m|\left(\frac{W_v}{4\,a}\right)^{4}\,\langle T\rangle_\varphi\,\delta(\hat{r}-\hat{r}_s),\label{e151}\\[0.5ex]
\frac{d\langle\hat{\varpi}_\theta\rangle_\varphi(0)}{d \hat{r}} =
\frac{d\langle\hat{\varpi}_\phi\rangle_\varphi(0)}{d \hat{r}} =0,\label{e152}\\[0.5ex]
\langle\hat{\varpi}_\theta\rangle_\varphi(1)=
\langle\hat{\varpi}_\phi\rangle_\varphi(1)=0.\label{e153}
\end{eqnarray}

\subsection{Solution of Cycle-Averaged Plasma Angular Equations of Motion}
Let \cite{chapman}
\begin{eqnarray}\label{e154}
\langle\hat{\varpi}_\theta\rangle_\varphi(\hat{r}) &=&\sum_{n=1,N} a_n\,\frac{y_n(\hat{r})}{y_n(\hat{r}_s)},\\[0.5ex]
\langle\hat{\varpi}_\phi\rangle_\varphi(\hat{r}) &=&\sum_{n=1,N} b_n\,\frac{z_n(\hat{r})}{z_n(\hat{r}_s)},\label{e155}
\end{eqnarray}
where
\begin{eqnarray}
y_n(\hat{r}) &=& \frac{J_1(j_{1,n}\,\hat{r})}{\hat{r}},\\[0.5ex]
z_n(\hat{r}) &=& J_0(j_{0,n}\,\hat{r}).\label{e158}
\end{eqnarray}
Here, $J_m(z)$ is a standard Bessel function, and $j_{m,n}$ denotes the $n$th zero of the $J_m(z)$ Bessel function \cite{abram1}.
Note that  (\ref{e154}) and (\ref{e155}) automatically  satisfy the boundary conditions (\ref{e152}) and (\ref{e153}), and are complete in the limit $N\rightarrow\infty$. 

It is easily demonstrated that \cite{abram1}
\begin{eqnarray}
\frac{d}{d\hat{r}}\!\left(\hat{r}^{\,3}\,\frac{dy_n}{d\hat{r}}\right)& =& - j_{1,n}^{\,2}\,\hat{r}^{\,3}\,y_n,\\[0.5ex]
\frac{d}{d\hat{r}}\!\left(\hat{r}\,\frac{dz_n}{d\hat{r}}\right) &=& - j_{0,n}^{\,2}\,\hat{r}\,z_n,
\end{eqnarray}
and\,\cite{grad2}
\begin{eqnarray}
\int_0^1\hat{r}^{\,3}\,y_n(\hat{r})\,y_m(\hat{r})\,d\hat{r}& = &\frac{1}{2}\left[J_2(j_{1,n})\right]^{\,2}\,\delta_{nm},\\[0.5ex]
\int_0^1\hat{r}\,z_n(\hat{r})\,z_m(\hat{r})\,d\hat{r}& = &\frac{1}{2}\left[J_1(j_{0,n})\right]^{\,2}\,\delta_{nm}\label{e162}
\end{eqnarray}
Thus, (\ref{e150})--(\ref{e151}) and (\ref{e154})--(\ref{e162}) yield 
\begin{equation}\label{etorque}
\langle\hat{\varpi}\rangle_\varphi = \hat{\varpi}_0-\sum_{n=1,N}a_n -\sum_{n=1,N}b_n,
\end{equation}
where
\begin{eqnarray}
a_n&=& \frac{\alpha_n}{1/\hat{\tau}_\theta+j_{1,n}^{\,2}/\hat{\tau}_M}\,|m|\left(\frac{W_v}{4\,a}\right)^{4}\,\langle T\rangle_\varphi,\\[0.5ex]
b_n&= &\frac{\zeta\,\beta_n}{1/\hat{\tau}_\phi+j_{0,n}^{\,2}/\hat{\tau}_M}\,|m|\left(\frac{W_v}{4\,a}\right)^{4}\,\langle T\rangle_\varphi,
\end{eqnarray}
and
\begin{eqnarray}
\alpha_n &= &\left[\frac{\sqrt{2}\,J_1(j_{1,n}\,\hat{r}_s)}{\hat{r}_s\,J_2(j_{1,n})}\right]^{2},\\[0.5ex]
\beta_n &= & \left[\frac{\sqrt{2}\,J_0(j_{0,n}\,\hat{r}_s)}{J_1(j_{0,n})}\right]^{2}.\label{e167}
\end{eqnarray}

\subsection{Cycle-Averaged Torque Balance}
Let
\begin{equation}
x=\frac{\langle \hat{\varpi}\rangle_\varphi}{\hat{\varpi}_0}.
\end{equation}
Equations~(\ref{e145}), (\ref{e149}), and (\ref{etorque})--(\ref{e167}) can be combined to give
the following cycle-averaged torque balance equation \cite{rfx}:
\begin{equation}
g(x)=0,\label{e169}
\end{equation}
where
\begin{equation}
g(x)\equiv \gamma\,x^{\,8/3}-\gamma\,x^{\,5/3} + \frac{\beta}{2\cdot 2^{\,1/3}}\,x-\frac{1}{2^{\,5}\cdot 2^{\,1/3}},
\end{equation}
and
\begin{eqnarray}
\beta&=& \frac{1}{2^{\,4}}+{\cal K}\,({\cal J}\,{\mit\Sigma})\left(\frac{{\cal I}\,S}{3}\right)\left(\frac{r_s}{a}\right)^4\left(\frac{W_v}{r_s}\right)^5,\\[0.5ex]
\gamma&=& {\cal K}\left(\frac{|\hat{\varpi}_0|}{|m|}\right)^{5/3}\left(\frac{{\cal I}\,S}{3}\right)^{5/3}\left(\frac{W_v}{r_s}\right)^{5/3},
\end{eqnarray}
 with
\begin{eqnarray}
{\cal J} &= &\frac{0.5356}{2^{\,6}},\\[0.5ex]
{\cal K} &=& \frac{0.8543}{2^{\,7}},\\[0.5ex]
{\mit\Sigma} &= &\sum_{n=1,N} \left(\frac{\alpha_n}{1/\hat{\tau}_\theta+j_{1,n}^{\,2}/\hat{\tau}_M}+\frac{\zeta\,\beta_n}{1/\hat{\tau}_\phi+j_{0,n}^{\,2}/\hat{\tau}_M}\right).
\end{eqnarray}
Let 
\begin{eqnarray}
\frac{W_c}{r_s} &=& \frac{(1-1/2^{\,4})^{1/5}}{{\cal K}^{\,1/5}\,({\cal J}\,{\mit\Sigma})^{\,1/5}\,({\cal I}\,S/3)^{1/5}
\,(r_s/a)^{4/5}},\\[0.5ex]
\frac{\hat{\varpi}_c}{|m|}& = &\frac{({\cal J}\,{\mit\Sigma})^{\,1/5}\,(r_s/a)^{4/5}}
{(1-1/2^{\,4})^{1/5}\,{\cal K}^{\,2/5}\,({\cal I} S/3)^{4/5}}.
\end{eqnarray}
It follows that
\begin{eqnarray}\label{e177}
\beta &=& \frac{1}{2^{\,4}}+\left(1-\frac{1}{2^{\,4}}\right)\left(\frac{W_v}{W_c}\right)^5,\\[0.5ex]
\gamma&=& \left(\frac{W_v}{W_c}\right)^{5/3}\left(\frac{|\hat{\varpi}_0|}{\hat{\varpi}_c}\right)^{5/3}.\label{e178}
\end{eqnarray}
Here, $x$ parameterizes the actual (i.e., in the presence of the RMP) plasma rotation at the
resonant surface, $\beta$ parameterizes the amplitude of the RMP, and $\gamma$ parameterizes
the intrinsic (i.e., in the absence of the RMP) plasma rotation at the resonant surface.

\begin{figure}
\includegraphics[height=4.5in]{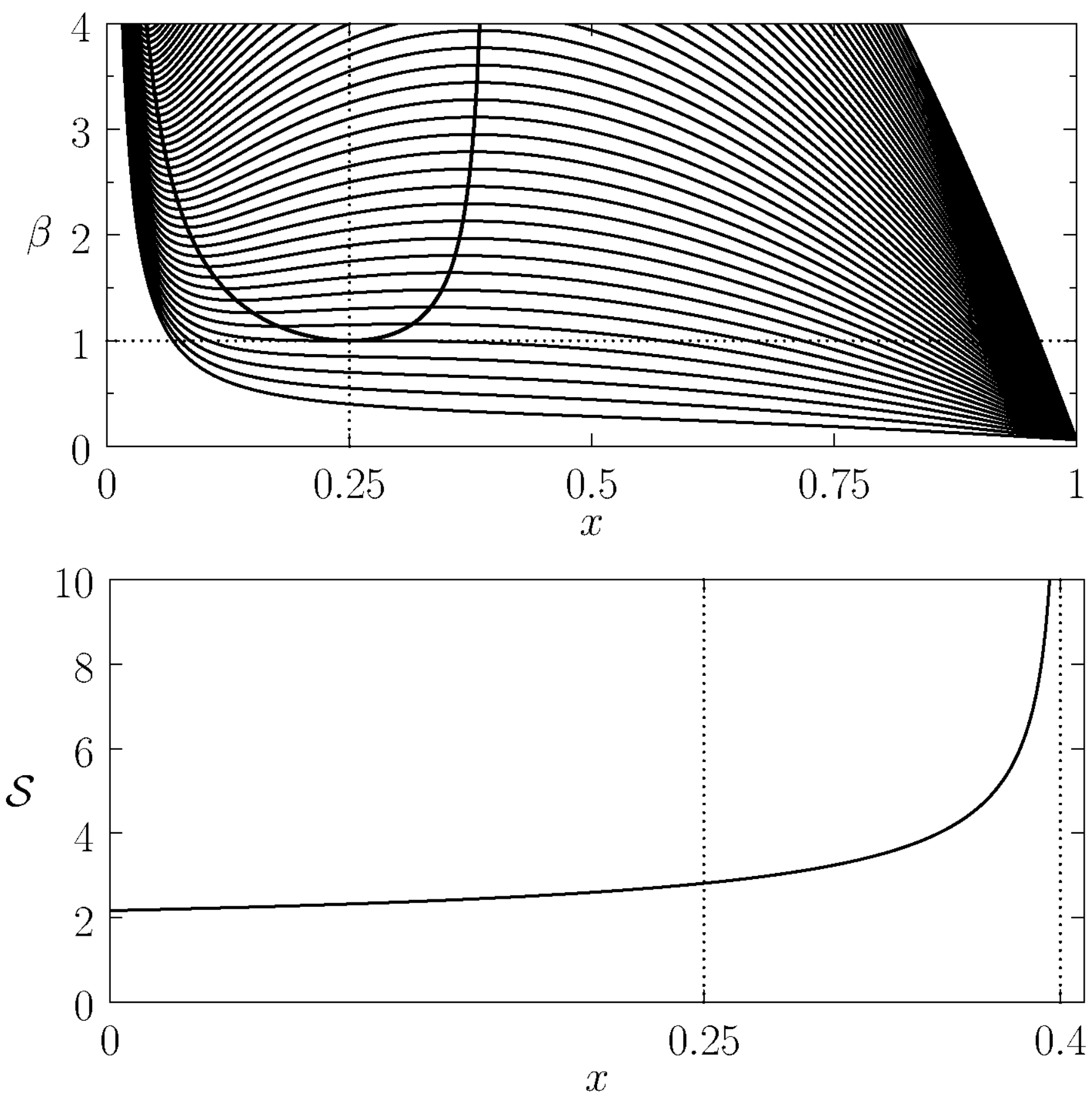}
\caption{Top Panel: Solutions of the cycle-averaged torque balance equation. The
thin curves show constant-$\gamma$ solutions plotted in $x$--$\beta$ space. The
curve that passes through the point $x=1/4$, $\beta=1$ corresponds to $\gamma=1$. Curves
that pass below (in $\beta$) this point correspond to $\gamma<1$, and vice versa. The
solutions lying within the thick curve are dynamically unstable.  Bottom Panel: The cycle-averaged shielding factor, ${\cal S}= 1/(\langle\hat{W}^{\,4}\rangle_\varphi)^{\,1/2}$, 
as a function of normalized plasma rotation at the resonant surface, $x$, for solutions of
the cycle-averaged torque balance equation that lie at the boundary between dynamically
stable and dynamically unstable solutions. Low-rotation solutions correspond to 
$0\leq x\leq 1/4$, whereas high-rotation solutions correspond to $1/4<x<2/5$.} \label{figa}
\end{figure}

\begin{figure}
\includegraphics[height=7in]{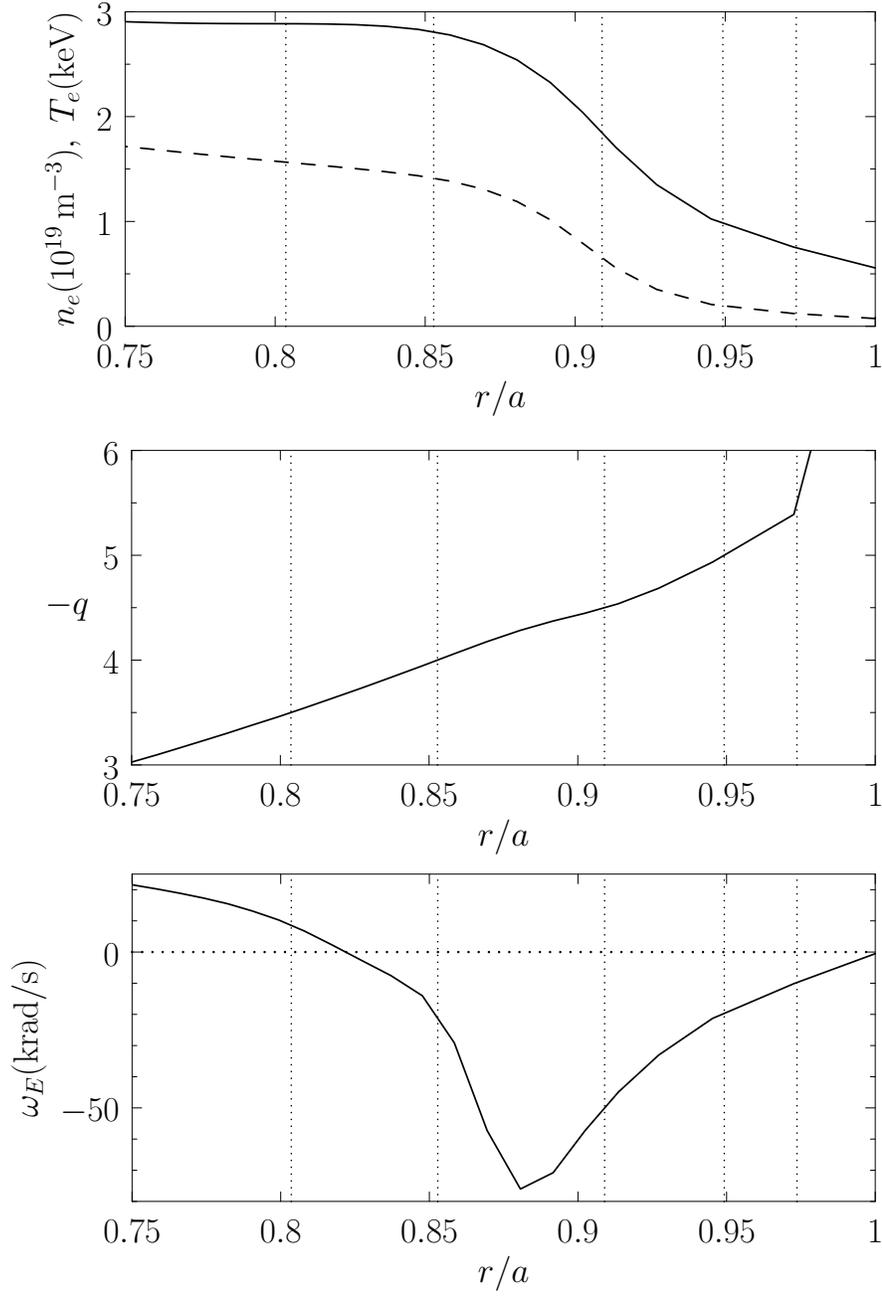}
\caption{Top Panel: Electron number density (solid curve) and electron temperature  (dashed curve) profiles in the pedestal region of 
DIII-D discharge \#158115 at time 3399 ms.  Middle Panel: Safety-factor profile in the pedestal region of 
DIII-D discharge \#158115 at time 3399 ms.  Bottom Panel: ${\bf E}\times {\bf B}$ frequency profile in the pedestal region of 
DIII-D discharge \#158115 at time 3399 ms.
Here, $r$ is the flux-surface averaged minor radius.  The
vertical dotted lines show the positions of the -7/2, -8/2, -9/2, -10/2, and -11/2 resonant surfaces, in order from left to right. Data reproduced  from Figure~2 of \cite{hu}.  } \label{figb}
\end{figure}

The numerical solution of the cycle-averaged torque balance equation, (\ref{e169}), is shown in Figure~\ref{figa} (upper panel). The solution exhibits a forbidden band of plasma rotation frequencies when $\gamma>1$
(i.e., when the intrinsic plasma rotation is sufficiently high). This band separates a branch
of dynamically stable, low-rotation solutions from a branch of dynamically stable, high-rotation
solutions. If a low-rotation solution crosses the lower boundary of the forbidden band then a
bifurcation to a high-rotation solution is triggered. Likewise, if a high-rotation
solution crosses the upper boundary of the forbidden band then a bifurcation to a low-rotation solution
is triggered. 

It is helpful to define the cycle-averaged shielding factor, 
\begin{equation}
{\cal S} =\frac{1}{(\langle\hat{W}^{\,4}\rangle_\varphi)^{\,1/2}}.
\end{equation}
This quantity represents the mean factor by which the reconnected magnetic flux driven at the
resonant surface is reduced by plasma rotation, relative to its value in the absence of
plasma rotation. Figure~\ref{figa} (lower panel) shows the shielding factor for solutions of the cycle-averaged
torque balance equation that lie on the boundary of the forbidden band of plasma rotation frequencies
(i.e., they lie on the thick curve in the upper panel of  Figure~\ref{figa}). It can be seen that low-rotation solutions are
characterized by weak shielding [i.e., ${\cal S}\sim {\cal O}(1)$], whereas high-rotation solutions
are characterized by strong shielding (i.e., ${\cal S}\gg 1$). Hence, high-rotation to low-rotation
bifurcations are associated with the sudden breakdown of strong shielding, whereas low-rotation
to high-rotation bifurcations are characterized by the sudden onset of strong shielding. 

To a good approximation, the bifurcation from the high-rotation to the
low-rotation solution branch occurs when $\beta$ exceeds the critical value \cite{rfx}
\begin{equation}\label{e180}
\beta_+\simeq  1+\frac{12}{5^{\,5/3}}\,(\gamma-1),
\end{equation}
whereas the bifurcation from the low-rotation to the high-rotation solution branch occurs when
$\beta$ falls below the critical value \cite{rfx}
\begin{equation}
\beta_-\simeq  1+\frac{5}{2\,(54)^{\,1/5}}\,(\gamma^{\,3/5}-1).
\end{equation}
The previous two expressions are only valid when $|\hat{\varpi}_0|>\hat{\varpi}_c$ (i.e, when $\gamma>1$). For $|\hat{\varpi}_0|\leq\hat{\varpi}_c$, there is
no forbidden band of plasma rotation frequencies, and, consequently, there are no bifurcations. 

Let the bifurcation from the 
high-rotation to the low-rotation solution branch occur when the  vacuum
island width exceeds the critical value $W_{v\,+}$. 
Furthermore, let 
\begin{equation}
y_{+}= \left(\frac{W_{v\,+}}{W_c}\right)^{5/3}.
\end{equation}
It follows from (\ref{e177}), (\ref{e178}), and (\ref{e180}) that $y_{+}$
is the most positive real root of 
\begin{equation}
\fl \left(1-\frac{1}{2^{\,4}}\right)y_+^{\,3}-\frac{12}{5^{\,5/3}}\left(\frac{|\hat{\varpi}_0|}{\hat{\varpi}_c}\right)^{5/3}y_+ + \frac{12}{5^{\,5/3}}-\left(1-\frac{1}{2^{\,4}}\right)=0.
\end{equation}
Likewise, let the bifurcation from the low-rotation to the high-rotation solution branch
occur when the vacuum island width falls below the critical value $W_{v-}$. 
Furthermore, let
\begin{equation}
y_{-}= \left(\frac{W_{v\,-}}{W_c}\right)^{5}.
\end{equation}
It follows that $y_{-}$
is the most positive real root of 
\begin{eqnarray}
\fl \left(1-\frac{1}{2^{\,4}}\right)y_{-}^{\,5}-\frac{5}{2\,(54)^{\,1/5}}\left(\frac{|\hat{\varpi}_0|}{\hat{\varpi}_c}\right)y_- 
+ \frac{5}{2\,(54)^{\,1/5}}-\left(1-\frac{1}{2^{\,4}}\right)=0.
\end{eqnarray}

Finally, let $b_v$ be the $m/n$ component of the vacuum radial magnetic field due to the RMP
at $r=a$. 
The bifurcation from the 
high-rotation to the low-rotation solution branch occurs when $b_v$ exceeds the critical value $b_{v\,+}$,
where 
\begin{equation}
\frac{b_{v\,+}}{|B_0|}=\frac{(n\,s)\,(a/R_0)\,y_+^{\,6/5}}{2^{\,4}\,(r_s/a)^{\,|m|-2}\,
\sqrt{1-\epsilon_s^{\,2}}
}\left(\frac{W_c}{r_s}\right)^2.
\end{equation}
[See (\ref{e98}), (\ref{e105}), and (\ref{e106}).] (Here, we are again assuming that there is
negligible equilibrium plasma current external to the resonant surface.) 
Likewise, the bifurcation from the low-rotation to the high-rotation solution branch occurs when
$b_v$ falls below the critical value $b_{v\,-}$, where 
\begin{equation}
\frac{b_{v\,-}}{|B_0|}=\frac{(n\,s)\,(a/R_0)\,y_-^{\,2/5}}{2^{\,4}\,(r_s/a)^{\,|m|-2}\,
\sqrt{1-\epsilon_s^{\,2}}
}\left(\frac{W_c}{r_s}\right)^2.
\end{equation}

\section{Locked Magnetic Island Chains}\label{sx}
\subsection{Steady-State Torque Balance}
A locked solution of (\ref{estart})--(\ref{eend})  (i.e., a solution in which the magnetic island chain driven at the resonant surface has a fixed phase relation with respect to the
RMP) is characterized by an island chain whose width and helical phase are constant in time.
The steady-state (i.e., $\partial/\partial \hat{t}=0$) versions of (\ref{estart})--(\ref{eend})
are: 
\begin{eqnarray}\label{e189}
\hat{W}=\cos^{1/2}\varphi,\\[0.5ex]
0 = \hat{\varpi}_0-\hat{\varpi}_\theta(\hat{r}_s)-\hat{\varpi}_\phi(\hat{r}_s),\\[0.5ex]
\fl -\frac{1}{\hat{\tau}_M}\,\frac{d}{d\hat{r}}\!\left(\hat{r}^{\,3}\,\frac{d\hat{\varpi}_\theta}{d\hat{r}}\right)+\frac{1}{\hat{\tau}_\theta}\,\hat{r}^{\,3}\,\hat{\varpi}_\theta
 = 
|m|\left(\frac{W_v}{4\,a}\right)^{4}\,\hat{W^{\,2}}\,\sin\varphi\,\delta(\hat{r}-\hat{r}_s),\\[0.5ex]
\fl \frac{1}{\hat{\tau}_M}\,\frac{d}{d\hat{r}}\!\left(\hat{r}\,\frac{d\hat{\varpi}_\phi}{d\hat{r}}\right)+\frac{1}{\hat{\tau}_\phi}\,\hat{r}\,\hat{\varpi}_\phi 
 = 
\zeta\,|m|\left(\frac{W_v}{4\,a}\right)^{4}\,\hat{W^{\,2}}\,\sin\varphi\,\delta(\hat{r}-\hat{r}_s),\\[0.5ex]
\frac{d\hat{\varpi}_\theta(0)}{d \hat{r}} =
\frac{d \hat{\varpi}_\phi(0)}{d\hat{r}} =0,\\[0.5ex]
\hat{\varpi}_\theta(1)=\hat{\varpi}_\phi(1)=0.
\end{eqnarray}
Making use of the representation (\ref{e154})--(\ref{e158}), we arrive at the following
steady-state torque balance equation:
\begin{equation}
\frac{\hat{\varpi}_0}{|m|}= \frac{{\mit\Sigma}}{2^{\,9}}\left(\frac{r_s}{a}\right)^4\left(\frac{W_v}{r_s}\right)^4\,\sin 2\varphi.
\end{equation}
Clearly, there is a bifurcation from the locked to the pulsating solution branch when $|\varphi|$ exceeds the critical value $\pi/4$, which corresponds to $W_v$ falling below the critical
value $W_{v\,0}$, where 
\begin{equation}
\left(\frac{W_{v\,0}}{r_s}\right)^2 = \frac{2^{\,9/2}\,(|\hat{\varpi}_0|/|m|)^{\,1/2}}{{\mit\Sigma}^{\,1/2}\,(r_s/a)^{\,2}}.
\end{equation}
Note, from (\ref{e189}), that the locked solution branch (for which $|\varphi|<\pi/4$) is not characterized by strong shielding [i.e., $\hat{W}\sim {\cal O}(1)$].
The bifurcation from the locked to the pulsating solution branch occurs when $b_v$ falls
below the critical value $b_{v\,0}$, where 
\begin{equation}
\frac{b_{v\,0}}{|B_0|} =\frac{(n\,s)\,(a/R_0)}{2^{\,4}\,(r_s/a)^{\,|m|-2}\,\sqrt{1-\epsilon_s^{\,2}}}\left(\frac{W_{v\,0}}{r_s}\right)^2.
\end{equation}

\subsection{Mode Penetration and Mode Unlocking Thresholds}
In most circumstances, $b_{v\,+}>b_{v\,0}> b_{v\,-}$, which implies that if $b_v$
exceeds the critical value $b_{v\,{\rm pen}}=b_{v\,+}$ then there is a bifurcation from the strongly-shielded, high-rotation,
pulsating solution branch to the weakly-shielded, locked solution branch. Moreover, if $b_v$ falls below the
critical value $b_{v\,{\rm ulk}}=b_{v\,0}$ then there is a bifurcation from the weakly-shielded, locked  solution branch
to the strongly-shielded, high-rotation, pulsating solution branch. In a relatively small number of cases,  $b_{v\,+}>b_{v\,-}> b_{v\,0}$, which implies that if $b_v$
exceeds the critical value $b_{v\,{\rm pen}}=b_{v\,+}$ then there is a bifurcation from the strongly-shielded, high-rotation, 
pulsating solution branch to the weakly-shielded,  low-rotation, pulsating solution branch. Moreover, if $b_v$ falls below the
critical value $b_{v\,{\rm ulk}}=b_{v\,-}$ then there is a bifurcation from the weakly-shielded, low-rotation,  pulsating solution branch
to the strongly-shielded, high-rotation, pulsating solution branch. Finally, if $|\hat{\varpi}_0|< \hat{\varpi}_c$ then 
$b_{v\,{\rm pen}}=b_{v\,{\rm ulk}}=b_{v\,0}$. In this case, there is a smooth transition from the 
pulsating solution branch to the locked solution branch when $b_v$ exceeds the critical value
$b_{v\,{\rm pen}}$, and a smooth transition  from the locked solution branch to the 
pulsating solution branch when  $b_v$ falls below the same critical value. 

\section{Application to DIII-D H-Mode Discharges}\label{sdiiid}
\subsection{DIII-D Discharge \#158115}
Following \cite{hu} and \cite{paper1}, this paper concentrates on a particular (but completely typical) DIII-D H-mode discharge (\#158115) \cite{d158115} in which
ELMs were successfully suppressed by an externally applied $n=2$ RMP. The electron number density, electron temperature, safety-factor, and ${\bf E}\times {\bf B}$
frequency profiles in the pedestal region of this discharge are shown in Figure~\ref{figb}  at a particular instance in time (3399 ms) at which the influence of the applied RMP on these profiles
is negligible. Note that, in this paper, the flux-surface label $r$ is identified with the flux-surface averaged minor radius. The remaining parameters that
characterize the discharge are $B_0=-1.94\,{\rm T}$, $R_0=1.75\,{\rm m}$, $a=0.93\,{\rm m}$, $Z_{\rm eff} = 2.5$, $M_i=2$, $M_I=12.011$  (which corresponds to
a deuterium plasma with carbon impurity ions), and
$\chi_\perp \equiv \rho_s/\mu_s=1\,{\rm m\,s}^{-2}$. Note that, in the following, the majority ion, impurity ion, and electron temperatures are all assumed to be equal, and toroidal flow-damping is neglected. 
According to \cite{hu}, the density pump-out is due to a locked magnetic island chain driven at the $m=-11/n=2$ resonant surface, which lies at
the bottom of the pedestal. Furthermore, ELM suppression is due to mode penetration at the $m=-8/n=2$ resonant surface, which lies at the top of the pedestal. 

\subsection{Edge Safety-Factor Scan}\label{sq1}
The most striking observed feature of RMP-induced ELM suppression  in an H-mode tokamak
discharge  is that it only occurs when $q_{95}$ takes values that lie in certain narrow
windows \cite{paz1,d158115}. In order to gain a better understanding of this phenomenon, we shall
crudely simulate a $q_{95}$-scan in DIII-D discharge \#158115 by taking the profiles shown in Figure~\ref{figb} and
adding a constant offset to the safety-factor: i.e., 
\begin{equation}\label{e199}
q(r)\rightarrow q(r)+{\mit\Delta}q.
\end{equation}
The unmodified equilibrium is characterized by $q_{95}=-4.34$. Varying the constant offset, 
${\mit\Delta} q$, allows us to change $q_{95}$, while keeping the density, temperature, and rotation
profiles constant. Moreover, by varying $q_{95}$, it is possible to move a given resonant surface all the
way though the pedestal region.  

\begin{figure}
\includegraphics[height=7in]{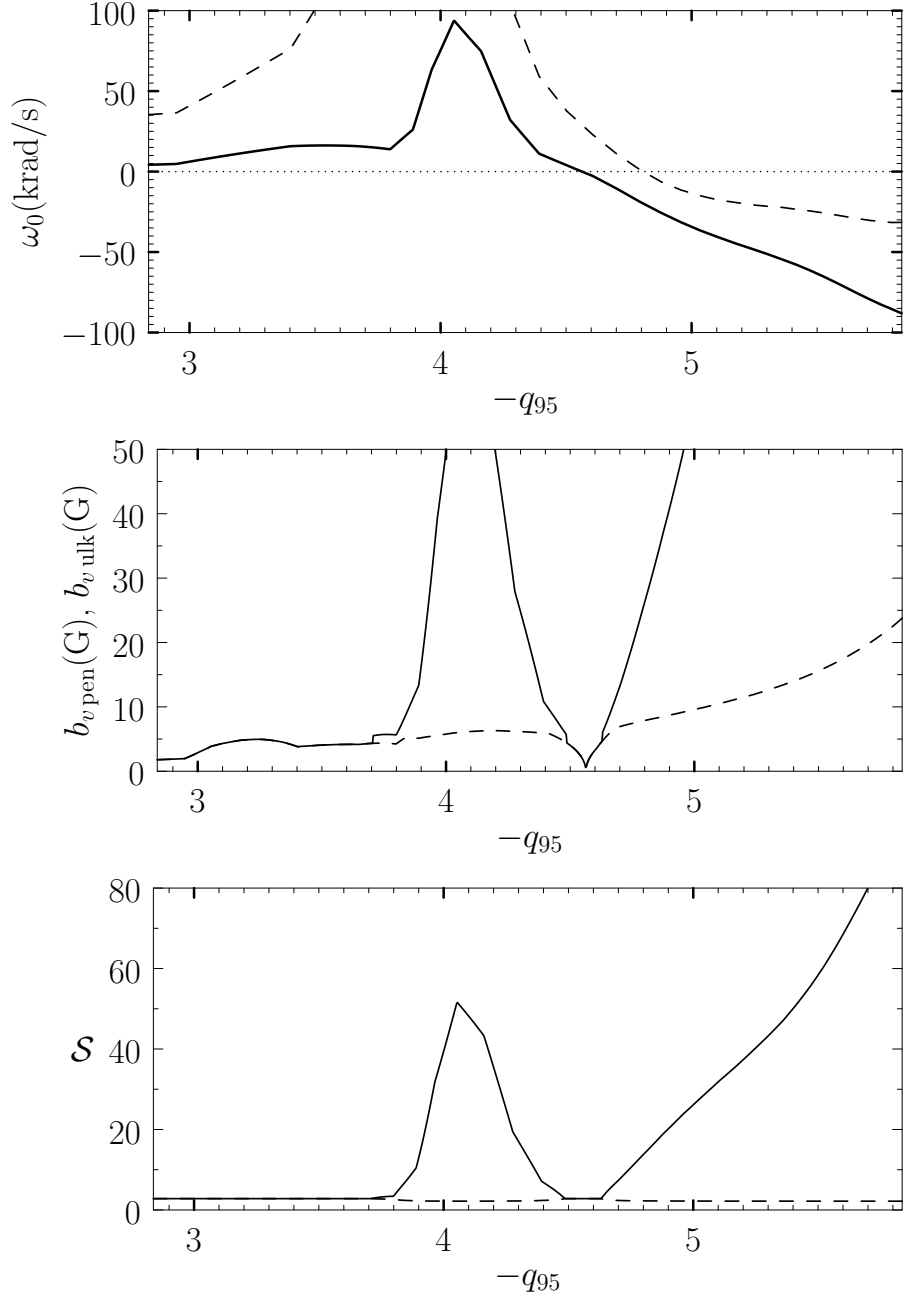}
\caption{Top Panel: Nonlinear (solid curve) and linear (dashed curve) natural phase velocities at  the $-8/2$ resonant surface in  DIII-D discharge \#158115 as functions of $q_{95}$.  Middle Panel: Mode penetration (solid curve) and unlocking (dashed curve) thresholds at the $-8/2$ resonant surface in  DIII-D discharge \#158115 as functions of $q_{95}$. Bottom Panel: Shielding factors just before mode penetration (solid curve) and mode unlocking (dashed curve) at the $-8/2$ resonant surface in  DIII-D discharge \#158115 as functions of $q_{95}$. } \label{figc}
\end{figure}

Figure~\ref{figc} specifies how the natural phase velocity, $\varpi_0$ [see (\ref{e123})], at the $m=-8/n=2$
resonant surface depends on $q_{95}$. For comparison, the
figure  shows the natural phase velocity predicted by linear theory, $\varpi_{\perp\,e} \equiv -n\,(\omega_E+\omega_{\ast\,e})_{r=r_s}$. It is clear  that $\varpi_0<\varpi_{\perp\,e}$. In other words, as expected, the nonlinear natural phase velocity is offset from the linear natural phase velocity in the ion diamagnetic (i.e., negative)  direction.
In addition, Figure~\ref{figc} specifies how the mode penetration and  unlocking thresholds at the $m=-8/n=2$
resonant surface, as well as the corresponding shielding factors just
before mode penetration and unlocking, depend on $q_{95}$. 

According to Figure~\ref{figc}, the plasma response at the $-8/2$ resonant surface is characterized by
strong shielding just prior to mode penetration   (i.e., ${\cal S}\gg 1$) and a large penetration threshold (i.e., $b_{v\,{\rm pen}}\gg 10\,{\rm G}$) for all values of $-q_{95}$ in the range $-q_{95}>3.8$, except for a narrow window of values
centered at $-q_{95}=4.55$. It should be noted that in DIII-D discharges the  maximum practical amplitude of the
vacuum radial magnetic field at the plasma boundary, $b_v$, for a particular helical harmonic of an applied
$n=2$ RMP, is typically about 10\,G \cite{hu}. Hence, if $b_{v\,{\rm pen}}\gg 10\,{\rm G}$ then mode
penetration at the $-8/2$ resonant surface is effectively impossible. 

Shielding breaks down when $-q_{95}<3.8$ because the $-8/2$
resonant surface is located in the cold resistive plasma at the bottom of the pedestal. Hence,
as discussed in Section~\ref{sadvance}, we would expect the application of a $-8/2$ RMP to the
plasma when $-q_{95}<3.8$ to give rise to a density pump-out, rather than ELM
suppression \cite{hu}. 
Shielding breaks down if $-q_{95}$ lies in a narrow window of
values centered on $-q_{95}=4.55$ because (as can be seen from Figure~\ref{figc}) the natural phase velocity passes through zero
when $-q_{95}=4.55$, so there is insufficient plasma ``rotation''  at the resonant surface to generate
strong shielding. Note that the $-8/2$ resonant surface lies close to the top of the pedestal when $-q_{95}=4.55$. As discussed in  Section~\ref{sadvance}, we would expect the application of a $-8/2$ RMP to the
plasma when $-q_{95}$ lies in the aforementioned narrow window of values to give rise to ELM
suppression \cite{hu}. On the other hand, we would expect the application of the RMP to the plasma when $-q_{95}$ exceeds 3.8,  but does not lie in the narrow window, to merely give rise to strong shielding. 
It can be seen, from Figure~\ref{figc}, that the width of the window of $-q_{95}$ values in which ELM
suppression is possible (which, roughly speaking, corresponds to the range of

values for which $b_{v\,{\rm pen}}< 10\,{\rm G}$) is about 0.3. This window width is less than the spacing (in $q$)
between the various resonant harmonics of the applied $n=2$ RMP (which is, of course, 0.5), implying that the $-q_{95}$ windows in  which the various resonant harmonics of the RMP
can cause ELM suppression in DIII-D discharge \#158115 do not overlap. 

According to \cite{paz}, there is a strong correlation between successful 
RMP-induced ELM suppression  in DIII-D plasmas and the ${\bf E}\times {\bf B}$ frequency, $\omega_E$,
passing through zero at a resonant surface located close to the top of the pedestal, but no correlation
whatsoever with the linear natural phase velocity, $\varpi_{\perp\,e}$, passing through zero at some such  surface. It turns out that, for the case illustrated in Figure~\ref{figc},  $-n\,\omega_E$ and the nonlinear natural phase velocity, $\varpi_0$, 
take very similar values when $-q_{95}\simeq 4.55$. Hence, the observations reported in \cite{paz}
are more consistent with the nonlinear mode penetration model presented in this paper than mode
penetration models that employ linear layer physics (which, inevitably, conclude that mode penetration
is correlated with $\varpi_{\perp\,e}$ passing through zero).

\subsection{Plasma Resistivity Dependance}
\begin{figure}
\includegraphics[height=7in]{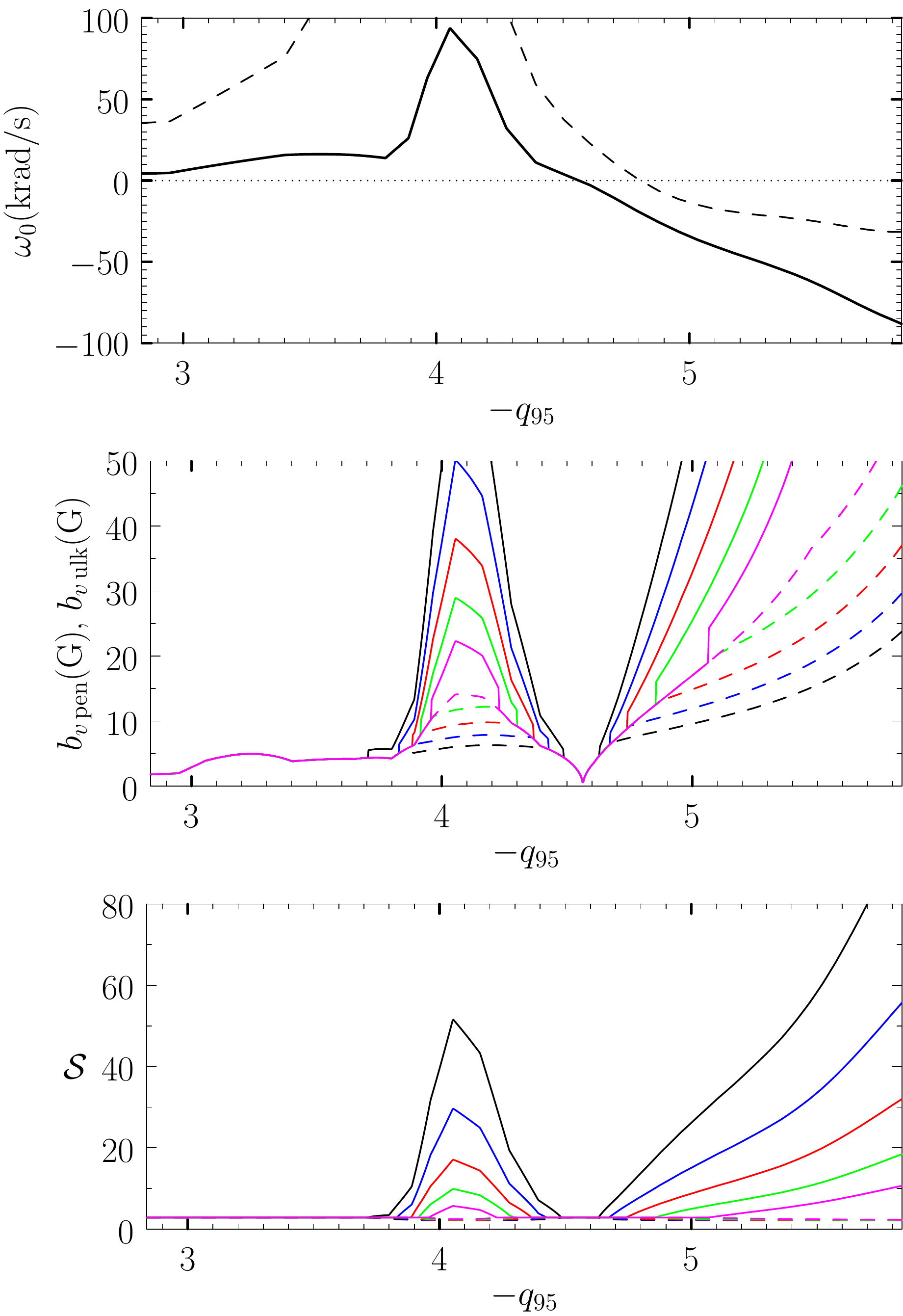}
\caption{Top Panel: Nonlinear (solid curve) and linear (dashed curve) natural phase velocities at  the $-8/2$ resonant surface in  DIII-D discharge \#158115 as functions of $q_{95}$.  Middle Panel: Mode penetration (solid curve) and unlocking (dashed curve) thresholds at the $-8/2$ resonant surface in  DIII-D discharge \#158115 as functions of $q_{95}$. The
  black, blue, red, green, and magneta curves correspond to calculations made with the experimental resistivity profile multiplied
  by factors of 1, 2, 4, 8, and 16, respectively. 
  Bottom Panel: Shielding factors just before mode penetration (solid curve) and mode unlocking (dashed curve) at the $-8/2$ resonant surface in  DIII-D discharge \#158115 as functions of $q_{95}$. The
  black, blue, red, green, and magneta curves correspond to calculations made with the experimental resistivity profile multiplied
  by factors of 1, 2, 4, 8, and 16, respectively.} \label{figca}
\end{figure}

Figure~\ref{figca} illustrates what happens to the calculation shown in Figure~\ref{figc} if we take the experimental
resistivity profile and scale it up by a constant factor. It can be seen that by the time the resistivity profile
has been increased by a factor 16 the strong shielding of driven magnetic reconnection in the interior of
the pedestal that is apparent in Figure~\ref{figc} has completely disappeared. This illustrates an important point.
Although it might be tempting to reduce the run time of a nonlinear-MHD simultation of RMP-induced ELM suppresssion
by increasing the plasma resistivity above its experimental value \cite{orain}, doing so risks completely changing the
nature of the plasma response to the RMP. If there is strong shielding in the interior of the pedestal, as we predict for DIII-D
discharge \#158115, then application of a multi-harmonic RMP with many resonant surfaces in the pedestal is only
likely to drive a magnetic island chain of significant width in the cold plasma at the bottom of the pedestal (where there
is no shielding) and, possibly, in the hot plasma at the top of the pedestal (where the natural phase velocity is particularly
small). Given that the two island chains are relatively far apart, the magnetic field in the pedestal is unlikely
to be rendered stochastic by the applied RMP. On the other hand, if there is no strong shielding in the interior
of the pedestal, as we predict if we increase the resistivity of DIII-D
discharge \#158115 by an order of magnitude, then application of a multi-harmonic RMP with many resonant surfaces in the pedestal is 
likely to drive  magnetic island chains of significant width  at  every resonant surface. In this case, it is almost inevitable that
the magnetic field in the pedestal will be rendered stochastic by the applied RMP. We note, finally, that
driving a single large magnetic island chain at the bottom of the pedestal causes a much larger reduction in the pedestal
density than in the pedestal temperature (because the density gradient at the bottom of the pedestal is larger than the
temperature gradient), which offers a plausible explanation of the density pump-out phenomenon \cite{hu}. On the
other hand, rendering the pedestal stochastic would inevitably cause a significant reduction in the pedestal temperature,
as well as the pedestal density, which cannot explain the density pump-out phenomenon. 

\subsection{Plasma Density Dependance}\label{sn}
\begin{figure}
\includegraphics[height=7in]{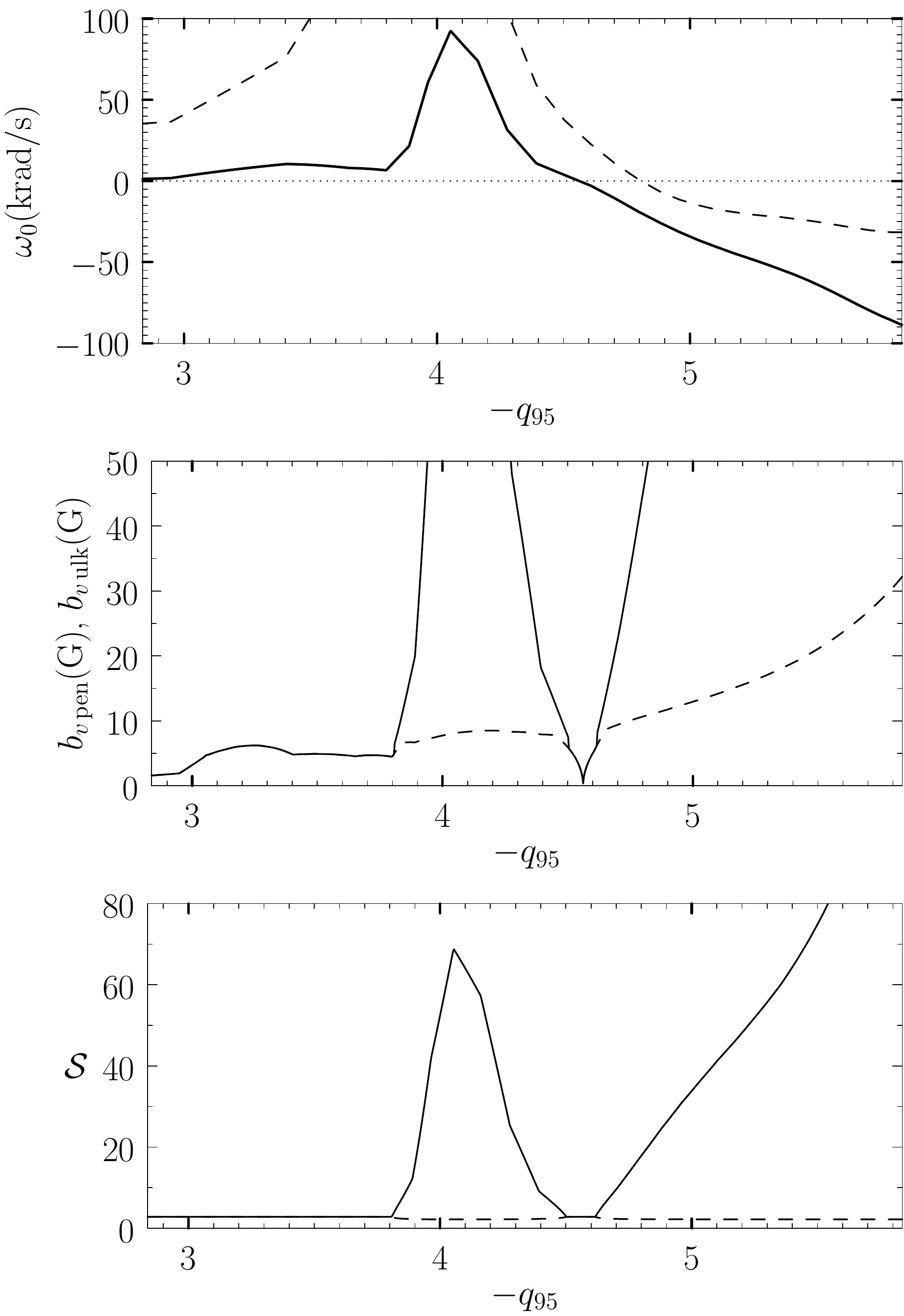}
\caption{Natural phase velocity, penetration and unlocking thresholds, and  shielding factors (see Figure~\ref{figc} caption) at the $-8/2$ resonant surface in DIII-D discharge \#158115 as
functions of $q_{95}$, with $n_e$ twice that specified in Figure~\ref{figb}.} \label{figd}
\end{figure}

RMP-induced ELM suppression is generally observed to be much more difficult when the  electron number density is comparatively high \cite{paz1,d158115}.
In order to gain a better understanding of this phenomenon, we shall crudely simulate  the effect of a density increase in DIII-D discharge \#158115
by taking the profiles shown in Figure~\ref{figb} and doubling the density: i.e.,
\begin{equation}
n_e(r)\rightarrow 2\,n_e(r).
\end{equation}
Figure~\ref{figd} illustrates the effect of such a density increase on the natural phase velocity, the mode penetration and unlocking thresholds, and the
shielding factors just before mode penetration and unlocking, at the $-8/2$ resonant surface. It can be seen, by comparison with Figure~\ref{figc}, that increasing the
plasma density does not strongly affect the natural phase velocity, but  leads to a large increase in the mode penetration threshold and the
associated shielding factor (except when the resonant surface lies in the cold plasma at the bottom of the pedestal), and, most importantly, produces a
significant shrinkage of  the width of the $q_{95}$-window in which RMP suppression is possible. Hence, the analytic model presented in this paper
is in general agreement with experimental observations regarding the influence of plasma density on RMP-induced ELM suppression. In fact, a careful examination
of the model reveals that increasing the plasma density increases both the plasma inertia and the ion collisionality. The increase in the plasma inertia leads to
a direct increase in the mode penetration threshold and the associated shielding factor, whereas the increase in the ion collisionality leads to
an indirect increase by increasing the ion neoclassical poloidal flow-damping rate. 

\subsection{Effect of High-$Z$ Impurities}\label{si1}
\begin{figure}
\includegraphics[height=7in]{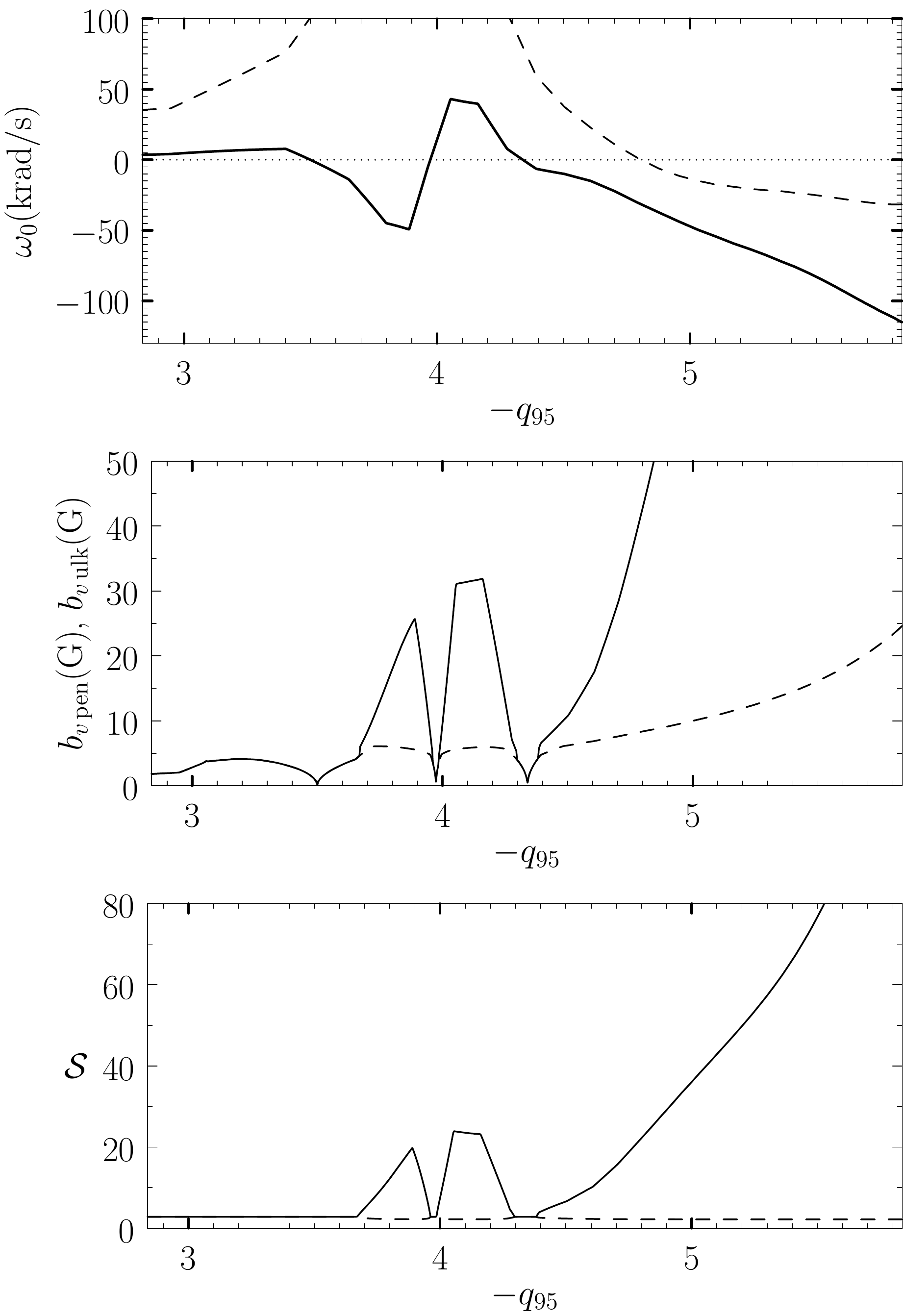}
\caption{ Natural phase velocity, penetration and unlocking thresholds, and  shielding factors (see Figure~\ref{figc} caption) at the $-8/2$ resonant surface in DIII-D discharge \#158115 as
functions of $q_{95}$, with tungsten, rather than carbon, as the impurity ion species. } \label{fige}
\end{figure}

\begin{figure}
\includegraphics[height=7in]{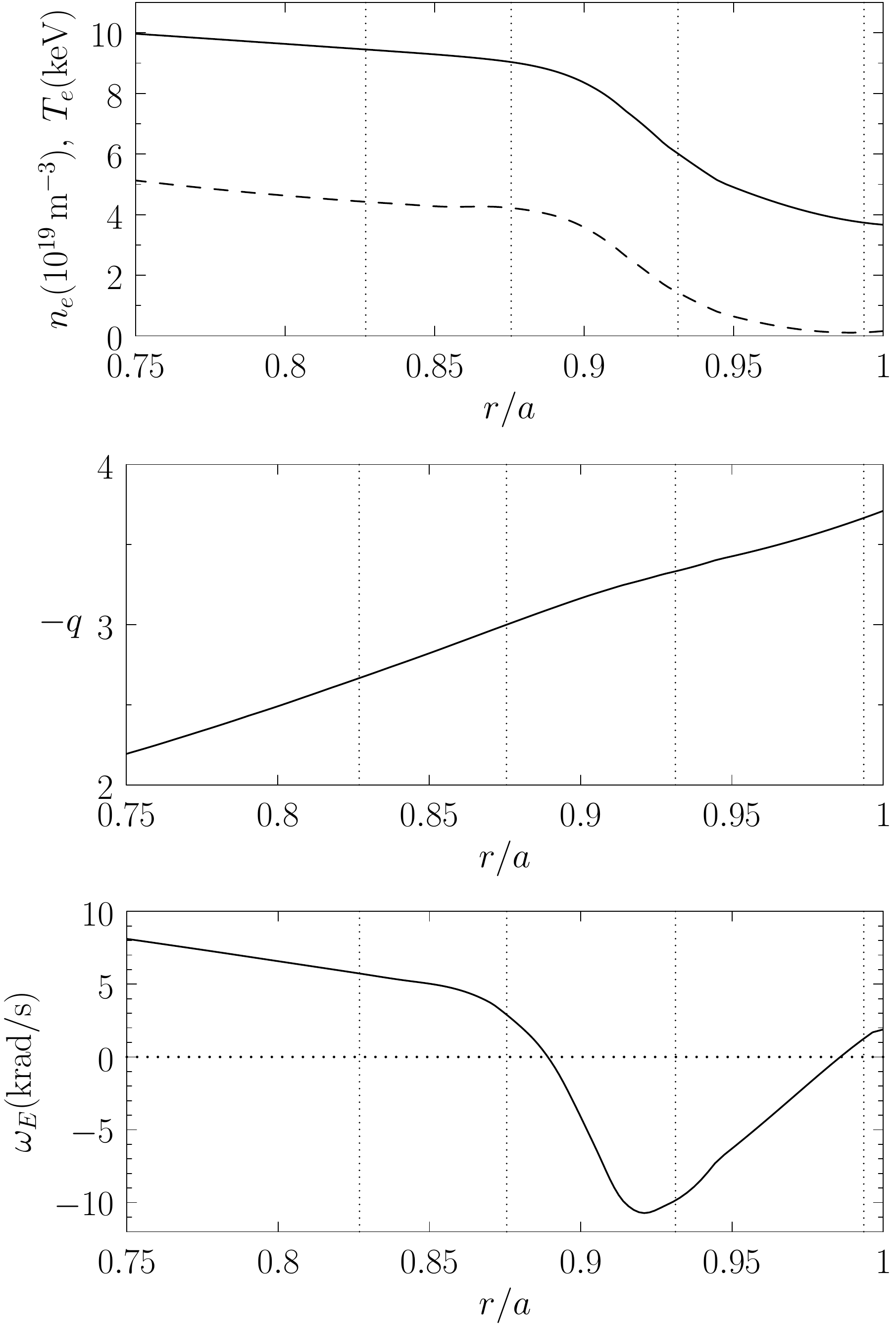}
\caption{Top Panel: Electron number density (solid curve) and electron temperature  (dashed curve) profiles in the pedestal region of 
a model ITER H-mode discharge.  Middle Panel: Safety-factor profile in the pedestal region of  a model 
 ITER H-mode discharge.  Bottom Panel: ${\bf E}\times {\bf B}$ frequency profile in the pedestal region of 
a model ITER H-mode discharge.
Here, $r$ is the flux-surface averaged minor radius.  The
vertical dotted lines show the positions of the -8/3, -9/3, -10/3, and -11/3 resonant surfaces, in order from left to right.  } \label{figf}
\end{figure}

Figure~\ref{fige} shows how the predictions of our analytic model of RMP-induced ELM suppression  for DIII-D discharge \#158115 are affected if the carbon impurities are replaced by tungsten impurities (keeping $Z_{\rm eff}$ the same). It can be seen that the replacement of carbon by tungsten as the main impurity species causes the natural phase velocity
at the $-8/2$ resonant surface to shift significantly in the ion diamagnetic (i.e., negative) direction. This shift presumably occurs because the tungsten ions are in the Pfirsch-Schl\"{u}ter collisionality
regime throughout the pedestal, whereas the carbon ions are in the banana regime at the top of the pedestal, and the plateau regime at the bottom. 
The aforementioned shift causes the opening of an additional very narrow $q_{95}$-window in which RMP-induced ELM suppression is possible.  The $-8/2$ resonant surface
is midway up the pedestal when $q_{95}$ lies in this window. The calculation described in this section illustrates the important role that plasma impurities can play in RMP-induced ELM
suppression. 

\section{Application to ITER H-Mode Discharges}\label{siter}
\subsection{Model ITER H-Mode Discharge}
Figure~\ref{figf} shows the electron number density, electron temperature, safety-factor, and ${\bf E}\times{\bf B}$ frequency profiles of a model ITER H-mode discharge. The profiles specified in the figure are
based on EPED \cite{eped1,eped2} and TGLF \cite{tglf} predictions \cite{men}. 
The rotation profile
is characterized by about 10 krad/s rotation close to the top of the pedestal, and about 20 krad/s
core rotation. This rotation profile is generated by a combination of an intrinsic torque and the torque
associated with 17 MW neutral beam injection power. The intrinsic torque and the momentum
confinement time are estimated from dimensionless parameter scans performed on the DIII-D
tokamak \cite{cry}.
The
remaining parameter that characterize the discharge are $B_0=-5.15$ T, $R_0=6.38$\,m, $a=1.98$\,m,
$Z_{\rm eff} = 1.6$, $M_i=2$, $M_I=9.1004$ (which corresponds to a deuterium plasma with
beryllium impurity ions), and $\chi_\perp = 1\,{\rm m\,s}^{-2}$.  (Note that we are assuming that the
ITER equilibrium magnetic field has the same helicity as the DIII-D field, for the sake of easy
comparison.) The majority ion, impurity ion, and electron temperatures are all assumed to be equal, and toroidal flow-damping is neglected. 

\subsection{Edge Safety-Factor Scan}\label{sq2}
As before, we shall crudely simulate a $q_{95}$-scan in our model ITER discharge by adding a constant offset to the
safety-factor profile [see (\ref{e199})], while keeping the other profiles fixed. The
unmodified equilibrium is characterized by $q_{95} =-3.08$. 

\begin{figure}
\includegraphics[height=7in]{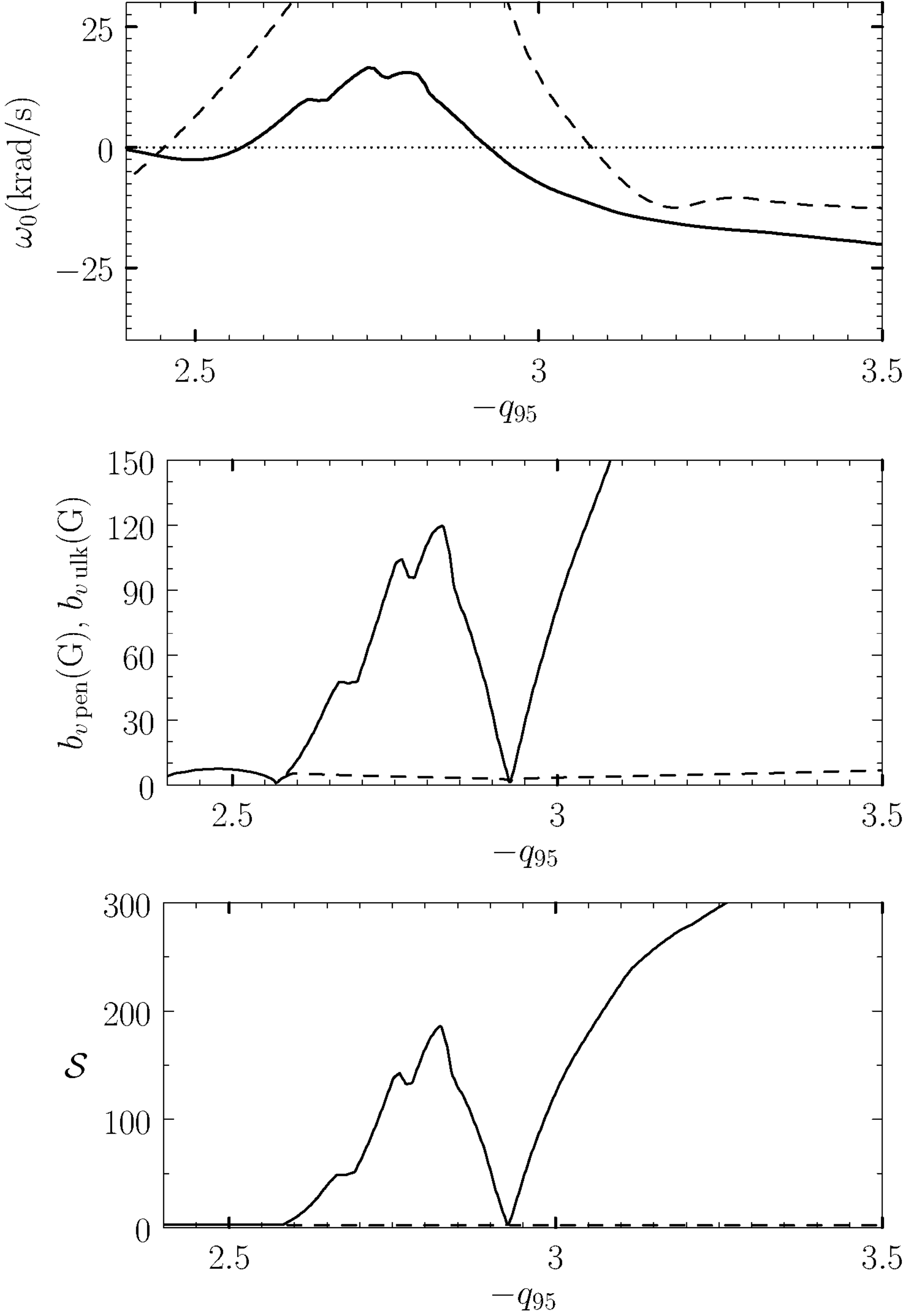}
\caption{Top Panel: Nonlinear (solid curve) and linear (dashed curve) natural phase velocities at  the $-9/3$ resonant surface  in  a model ITER H-mode discharge as functions of $q_{95}$.  Middle Panel: Mode penetration (solid curve) and unlocking (dashed curve) thresholds at the $-9/3$ resonant surface in  a model ITER H-mode discharge as functions of $q_{95}$. Bottom Panel: Shielding factors just before mode penetration (solid curve) and mode unlocking (dashed curve) at the $-9/3$ resonant surface in  a model ITER H-mode discharge  as functions of $q_{95}$.} \label{figg}
\end{figure}

Figure~\ref{figg} specifies how the natural phase velocity, $\varpi_0$ [see (\ref{e123})], at the $m=-9/n=3$
resonant surface depends on $q_{95}$. For comparison, the
figure  shows the natural phase velocity predicted by linear theory, $\varpi_{\perp\,e} \equiv -n\,(\omega_E+\omega_{\ast\,e})_{r=r_s}$. As expected, the nonlinear natural phase velocity is offset from the linear natural phase velocity in the ion diamagnetic   direction.
In addition, Figure~\ref{figg} specifies how the mode penetration and  unlocking thresholds at the $m=-9/n=3$
resonant surface, as well as the corresponding shielding factors just
before mode penetration and unlocking, depend on $q_{95}$. 

The data shown in Figure~\ref{figg} is qualitatively similar to that shown in Figure~\ref{figc}. 
Shielding at the $-9/3$ resonant surface breaks down for $-q_{95}< 2.6$ because the surface
is located in the cold resistive plasma at the bottom of the pedestal. Hence, we would expect the application of a $-9/3$ RMP to the
plasma when $-q_{95}<2.6$ to give rise to a density pump-out, rather than ELM
suppression \cite{hu}. The data shown in the figure indicates 
strong shielding just prior to mode penetration   (i.e., ${\cal S}\gg 1$), combined with a large penetration threshold (i.e., $b_{v\,{\rm pen}}\gg 60\,{\rm G}$), for all values of $-q_{95}$ in the range $-q_{95}>2.6$, except for a narrow window of values
centered at $-q_{95}=2.93$. Shielding breaks down if $-q_{95}$ lies in this narrow window of values because (as can be seen from the figure) the natural phase velocity passes through zero
when $-q_{95}=2.93$, so there is insufficient plasma ``rotation''  at the resonant surface to generate
strong shielding. As before, the $-9/3$ resonant surface lies close to the top of the pedestal when $-q_{95}=2.93$.  We would expect the application of a $-9/3$ RMP to the
plasma when $-q_{95}$ lies in the aforementioned narrow window of values to give rise to ELM
suppression \cite{hu}. On the other hand, we would expect the application of the RMP to the plasma when $-q_{95}$ exceeds 2.6,  but does not lie in the narrow window, to merely give rise to strong shielding. 

The ITER ELM control system is designed to have twice the 
 relative (to the toroidal magnetic field strength) capability of the present DIII-D system \cite{evansx,evansy}. 
Hence, we infer that the maximum practical amplitude of the
vacuum radial magnetic field at the plasma boundary, $b_v$, for a particular helical harmonic of an applied
$n=3$ RMP, is typically about 60\,G.  It follows from Figure~\ref{figg} that the width of the window of $q_{95}$ values in which $n=3$ ELM
suppression is possible in ITER  (which, roughly speaking, corresponds to the range of
values for which $b_{v\,{\rm pen}}< 60\,{\rm G}$) is about 0.1. This is slightly smaller
than the window width for effective ELM control in DIII-D with an $n=2$ RMP that we found in 
Section~\ref{sdiiid}. 

\subsection{Effect of High-$Z$ Impurities}\label{si2}
\begin{figure}
\includegraphics[height=7in]{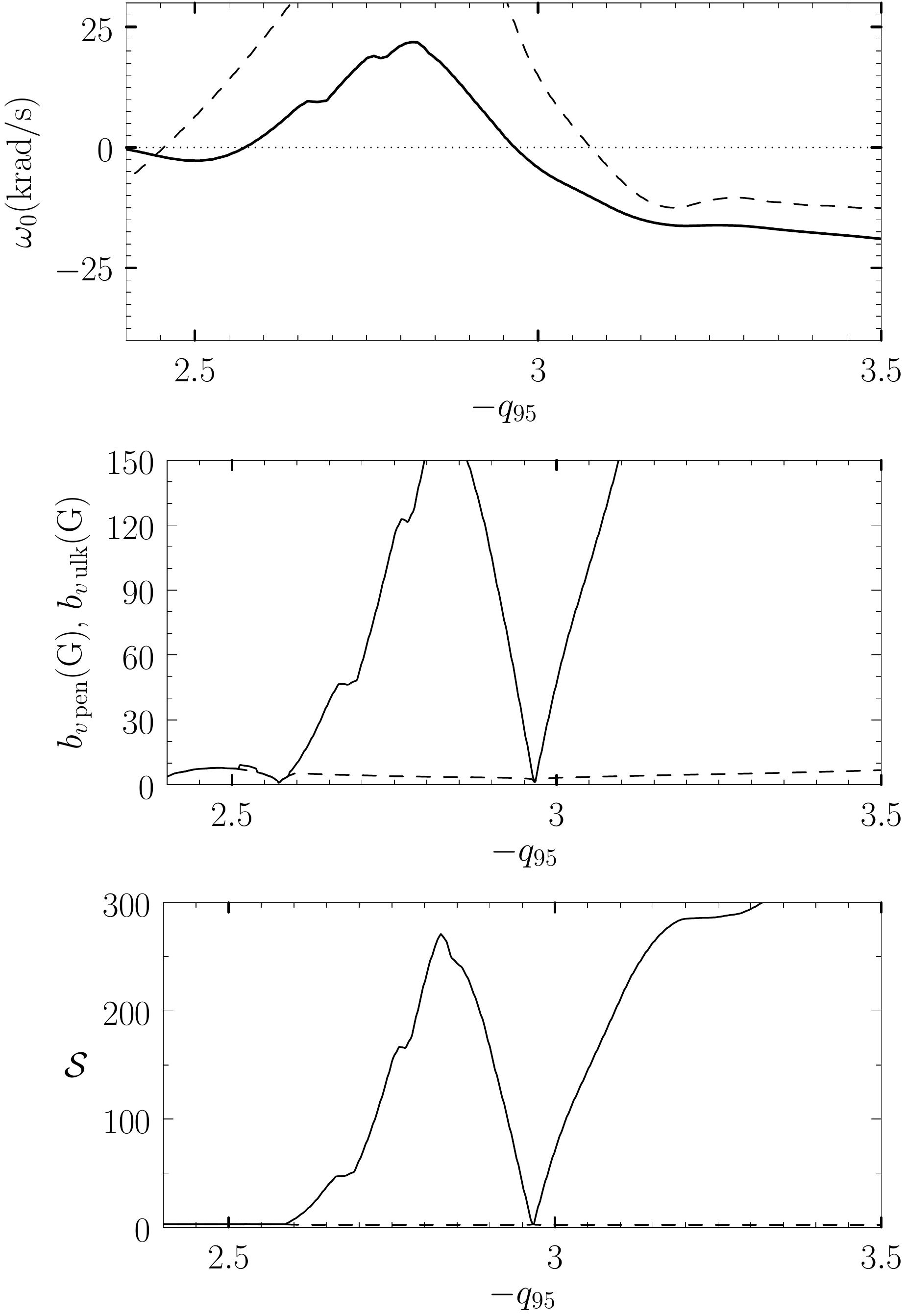}
\caption{Natural phase velocity, penetration and unlocking thresholds, and  shielding factors (see Figure~\ref{figg} caption) at  the $-9/3$ resonant surface  in  a model ITER H-mode discharge as functions of $q_{95}$,  with tungsten  as the impurity ion species.} \label{figh}
\end{figure}

Figure~\ref{figh} shows how the predictions of our analytic model of RMP-induced ELM suppression  for a
model ITER H-mode discharge are affected if the beryllium impurities are replaced by tungsten impurities (keeping $Z_{\rm eff}$ the same). It can be seen that the replacement of beryllium by tungsten as
the main impurity species causes a subtle modification in the natural phase velocity profile which, in turn, 
causes the window of $q_{95}$ values in which ELM suppression is possible to shift position slightly. 
Given that successful RMP-induced ELM control in ITER essentially boils down to locating the
aforementioned window, the calculation presented in this section again highlights the important
role that impurities play in the theory of ELM suppression.

\section{Summary and Conclusions}
We have developed an analytic theory of ELM suppression in an H-mode tokamak plasma via the application of a static, externally generated, RMP.  (See Sections~\ref{s2}--\ref{sx}.) 
This theory is based on the plausible hypothesis that mode penetration at the top of the pedestal is a necessary and
sufficient criterion for RMP-induced ELM suppression \cite{hu}. 
The theory also makes use of
  a number of key insights obtained in \cite{paper1}. The first insight is that the response of the plasma to a particular helical component of the RMP, in the immediate vicinity of the associated resonant surface, is governed by nonlinear magnetic island physics, rather than by linear layer physics. This has two important consequences. First, the shielded state consists of a rotating, pulsating island chain, rather than a stationary island chain. (See Section~\ref{spul}.)  Second, the so-called natural phase velocity of the driven island chain (i.e., its preferred phase velocity in the absence of the RMP) is quite different to that predicted by linear theory. According to linear layer theory, a naturally unstable island chain propagates in the electron diamagnetic direction relative to the local guiding center frame. However, according to nonlinear island theory, the island chain propagates in the ion diamagnetic direction relative to the local guiding center frame. (See Section~\ref{snat}.) This is significant because both the degree of shielding and the mode penetration threshold at the resonant surface depend crucially on the natural phase velocity. The second insight is that neoclassical effects play a vital role in the physics of mode penetration. This is the case, firstly, because intrinsic neoclassical poloidal rotation is the main controlling factor that determines the natural phase velocity  of a nonlinear magnetic island chain (see Section~\ref{snat}), and,
secondly, because neoclassical poloidal flow-damping plays a very significant role in determining the mode penetration threshold. The final insight is that plasma impurities play an important role in the physics of mode penetration. This is the case because impurities can significantly modify both the intrinsic neoclassical poloidal rotation and the neoclassical poloidal flow-damping rate.  

The theory developed in this paper  has been used
 to gain a better understanding of the physics of RMP-induced ELM suppression in DIII-D and ITER H-mode discharges.  (See Sections~\ref{sdiiid} and \ref{siter}.) It is found that ELM suppression
 is only possible when $q_{95}$  takes values that lie in certain narrow windows. (See Sections~\ref{sq1} and \ref{sq2}.)  Moreover, the widths of these windows decrease with increasing
 plasma density. (See Section~\ref{sn}.) The location and widths of the windows are also sensitive functions of both the concentration and the masses of the plasma
 impurities. (See Section~\ref{si1} and \ref{si2}.) 
  The window width for a model ITER H-mode discharge is found to be similar to, but slightly smaller than, the window width in a typical DIII-D
 H-mode discharge. 

There are a number
of improvements that could be made to the theory described in this paper.  Such improvements include; 1) taking into account the  coupling
of different resonant surfaces via mode-penetration-induced changes  in the plasma rotation, density, and temperature, profiles  in the pedestal \cite{hu}; 2)
taking into account the coupling of different poloidal harmonics of the RMP due to
toroidicity, the Shafranov shift, and flux surface shaping \cite{tor}; 3) employing a more realistic plasma equilibrium; 4)
including island saturation terms \cite{sat1,sat2,sat3},
the perturbed bootstrap current \cite{hel},  and the perturbed ion
polarization current \cite{ionpz}, in the Rutherford equation; 5) taking into account orbit-squeezing effects due to the strong shear in the
radial electric field that is typically present in H-mode tokamak pedestals \cite{orbit}; 6) taking neoclassical toroidal flow-damping into account; 7)
allowing for multiple impurity species; 
and 8) taking into account the fact that high-Z impurities
can easily acquire supersonic neoclassical velocities \cite{super}.

\section*{Acknowledgements}
This research was funded by the U.S.\ Department of Energy under contract DE-FG02-04ER-54742.
The author would like to thank Q.M.~Hu, R.~Nazikian, C.~Paz-Soldan, T.E.~Evans,  C.~Chrystal, and R.D.~Hazeltine for
helpful discussions.

\newpage

\end{document}